\documentclass[prd,nobibnotes,nofootinbib,onecolumn,10pt,a4paper,showpacs]{revtex4}

\usepackage{graphicx}
\usepackage{amssymb,latexsym}
\usepackage{varioref,xr-hyper}
\usepackage[draft=false]{hyperref}
\usepackage[sort&compress]{natbib}
\usepackage{hypernat}

\begin{document}

\title{Scalar perturbations in an $\alpha'$-regularised cosmological bounce}

\author{\firstname{Cyril} \surname{Cartier}}
\email[E-mail: ]{cyril.cartier@physics.unige.ch}
\affiliation{Section de Physique, Universit\'{e} de Gen\`{e}ve,\\
             Quai E. Ansermet 24, 1204 Gen\`{e}ve, Switzerland}

\date{January 7, 2004}

\begin{abstract}
  We consider the evolution of scalar perturbations in a class of
  non-singular bouncing universes obtained with higher-order
  corrections to the low-energy bosonic string action. We show that
  previous studies have relied on a singular evolution equation for
  the perturbations. From a simple criterium we show that scalar
  perturbations cannot be described at all times by an homogeneous
  second-order perturbation equation in pre-big bang type universes if
  we are to regularise the background evolution with higher-order
  curvature and string coupling corrections, and we propose a new
  system of first-order coupled differential equations. Given a
  bouncing cosmological background with inflation driven by the
  kinetic energy of the dilaton field, we obtain numerically the final
  power spectra generated from the vacuum quantum fluctuations of the
  metric and the dilaton field during inflation. Our result shows that
  both Bardeen's potential, $\Phi(\eta,k)$, and the curvature
  perturbation in the uniform curvature gauge, ${\mathcal R}(\eta,k)$,
  lead to a blue spectral distribution long after the transition.
\end{abstract}

\preprint{hep-th/0401036}
\maketitle

\section{Introduction}
\label{s:introduction}

Although current
observations~\cite{Bennett:2003bz,Spergel:2003cb,Hinshaw:2003ex}
of anisotropies in the cosmic microwave background radiation
(CMBR) strongly support the paradigm of scalar field
potential-driven inflation as being the source of primordial
density fluctuations, leading to an almost flat power
spectrum~\cite{Book_Inflation_Kolb,Book_Inflation_Linde}, this cosmological
model is not short of conceptual
problems~\cite{Brandenberger:1999sw}. One such loophole is the
initial singularity which persists in standard
inflation~\cite{Borde:1994xh}, but may eventually be addressed
within a more fundamental theory of quantum gravity. String theory
for instance admits a class of cosmological models where a period
of super-inflation is driven by the kinetic energy of a scalar
field. This ``pre-big bang'' phase with growing coupling and
growing curvature is expected to last until the background
evolution reaches a regime of maximal curvature, the bounce, where
pure string effects may eventually turn the evolution to an
expanding Friedmann-Lema\^{i}tre (FL) regime. To represent a
viable alternative to standard inflation, these cosmological
models should also reproduce the high-accuracy measurements of the
CMBR. To date, the late time spectral distribution of adiabatic
density perturbations in pre-big bang type universes remains the
subject of an intense debate. Some
authors~\cite{Khoury:2001wf,Khoury:2001zk,Durrer:2002jn,Peter:2002cn}
recently argued that a scale invariant spectrum (for the ekpyrotic
model~\cite{Khoury:2001wf,Khoury:2001zk}) or a red spectrum (for
the pre-big bang
scenario~\cite{Veneziano:1991ek,Gasperini:1993em}) might emerge
from the bouncing regime, while
others~\cite{Brandenberger:2001bs,Lyth:2001pf,Hwang:2001ga}
favoured a steep blue spectrum with $n\simeq 3$ (or $n=4$,
respectively). It is thus of prime interest to determine precisely
the spectral distribution of cosmological models which transit
from a collapsing phase to an expanding FL regime. Since this
discrepancy in the final spectral distribution results from
applying different matching procedures to tree-level singular
models, a regularised background evolution could in principle
provide the framework to discriminate between these opposite
claims.

One possibility of smoothing out a curvature singularity of the
background evolution is to enhance the low-energy effective action
of string theory with higher-order curvature corrections (e.g.,
see~\cite{Antoniadis:1994jc,Brustein:1998cv,Foffa:1999dv,Cartier:1999vk}).
The general feature that emerges from this scheme is that
higher-order corrections generally saturate the growth of
spacetime curvature, while quantum loop corrections are required
to violate the null energy condition and trigger a graceful exit
to the post-bounce decelerating phase. The inclusion of
higher-order corrections in the perturbation equations eventually
modify the evolution of adiabatic density
perturbations~\cite{Cartier:2001is}. Lately, these perturbation
equations have been used~\cite{Tsujikawa:2002qc} to study the
ekpyrotic model. In this work, the authors have followed the
perturbations through a regularised transition and have obtained a
final $n\simeq 3$ spectrum. Here we shall argue that their method
relied on a singular perturbation equation. We will then propose a
new homogeneous system of coupled linear differential equations
involving Bardeen's potential $\Phi$ and curvature perturbation in
the comoving gauge, ${\cal R}$. This system remains well-defined
at all times. Numerical simulations then provide us with
post-bounce spectral indices which we find are identical,
$n_{\Phi} = n_{\cal R} \simeq 3$ (or $n_{\Phi} = n_{\cal R} =4$
for the pre-big bang scenario). Thus our result appears in
contradiction with the spectral index advocated by Durrer and
Vernizzi~\cite{Durrer:2002jn} and favours the steep blue spectrum
put forward in~\cite{Finelli:2001sr} and others.

The paper is organised as follows. In
Sec.~\ref{s:action-GenGravity} we introduce a general action
including possible curvature corrections up to fourth order in
derivatives and we derive the corresponding field equations.
Section~\ref{s:G-perturbations} is devoted to studying the
classical evolution of adiabatic density perturbations. We first
comment on previous results and demonstrate that, in our
particular class of cosmologies, the perturbation equations cannot
be reduced to a set of one decoupled homogeneous second order
differential equation for each variable. Given a background
evolution, we then confront our predictions for the late time
spectral distributions by numerically integrating a new set of
perturbations equations. Finally, in Sec.~\ref{s:conclusion} we
recall our main results and discuss possible extensions. For the
sake of clarity, we leave the details of the calculation to the
Appendix.


\section{Background evolution}
\label{s:action-GenGravity}

As a starting point, we consider the following $D$-dimensional
effective action
\begin{equation}
{\cal S} = \frac{1}{\ell^{D-2}} \int d^D x
           \sqrt{-g} \left[ \frac{1}{2} {\cal F}(\phi) R
          -\frac{1}{2} \omega (\phi) \phi^{;\mu} \phi_{;\mu}
          - {\cal V} (\phi)
          + \frac{1}{2}\alpha'\lambda {\cal L}^{(c)}\right]~,
          \label{e:general-action}
\end{equation}
with $\ell$ a spacetime-length parameter and the
higher-order correction terms are given by
\begin{equation}
   {\cal L}^{(c)} \equiv \xi(\phi) \left[ c_1 R_{GB}^2
   + c_2 G^{\mu\nu} \phi_{;\mu} \phi_{;\nu}
   + c_3 \Box \phi \phi^{;\mu} \phi_{;\mu}
   + c_4 (\phi^{;\mu} \phi_{;\mu})^2 \right]~,
   \label{e:action-corr-gen}
\end{equation}
where $R_{GB}^2=R_{\mu\nu\lambda\rho}^2-4R_{\mu\nu}^2+R^2$ is the
Gauss-Bonnet combination. Here ${\cal F}(\phi)$, $\omega(\phi)$
and $\xi(\phi)$ are algebraic functions of a dimensionless scalar
field $\phi$, and with the potential ${\cal V}(\phi)$ we leave
open the possibility of scalar field self-interactions. Through
the lagrangian density ${\cal L}^{(c)}$, we allow for the
inclusion of terms with higher numbers of derivatives such as
contracted quadratic products of the curvature tensor and define
its associated energy-momentum tensor by
$\frac{1}{2}\sqrt{-g}~T^{(c)\,\mu\nu} \delta g_{\mu\nu} \equiv
\delta (\sqrt{-g}~{\cal L}^{(c)})$. Extremising the general
effective action Eq.~(\ref{e:general-action}) yields the covariant
Euler-Lagrange equations for the cosmological model,
\begin{eqnarray}
 && G^\mu_\nu = \frac{1}{2{\cal F}}\left[{T^{(0)}}^\mu_\nu
   + \alpha'\lambda {T^{(c)}}^\mu_\nu\right]~,
   \label{e:G-def-Einstein-eff}\\
 && 2\omega \Box \phi + {\cal F}_{,\phi}R
   +\omega_{,\phi}\phi^{;\mu}\phi_{;\mu}-2{\cal V}_{,\phi}
   + \alpha'\lambda \Delta^{(c)}_\phi = 0~,
   \label{e:G-EOM-phi}
\end{eqnarray}
where ${T^{(0)}}^\mu_\nu$ is the energy-momentum tensor of the
tree-level terms. The variation of the lagrangian density
Eq.~(\ref{e:action-corr-gen}) with respect to the scalar field
$\phi$ yields the $\alpha'$ correction $\Delta^{(c)}_\phi$ which
satisfies $ {T^{(c)}}^\mu_{\nu;\mu} = \Delta^{(c)}_\phi
\phi_{;\nu}$ according to Bianchi's identity. Explicitly, they are
given by~\cite{Cartier:2001is}
\begin{eqnarray}
 {T^{(0)}}^\mu_\nu &\equiv &
 2\left(\omega+{\cal F}_{,\phi\phi}\right)
 \phi^{;\mu}\phi_{;\nu}+2{\cal F}_{,\phi}{\phi^{;\mu}}_\nu
 -\delta^\mu_\nu\big[\left(\omega+2{\cal F}_{,\phi\phi}\right)
 \phi^{;\sigma}\phi_{;\sigma}+2{\cal F}_{,\phi}\Box\phi+2{\cal V}(\phi)\big]~,
 \label{e:G-def-T0} \\
 && \nonumber \\
   {T^{(c)}}^\mu_\nu &\equiv&
    - 8 c_1 \left[ \left({R^{\mu}}_{\sigma\nu\tau} +R_{\nu\sigma}
    \delta^{\mu}_{\tau}-\delta^{\mu}_{\nu}R_{\sigma\tau}
    \right)\xi^{;\sigma\tau}
    +G^{\mu\sigma}\xi_{;\nu\sigma}-G^{\mu}_{\nu}\Box\xi\right]+
    2c_1 \xi \aleph^\mu_\nu
 \nonumber \\ &&
   + c_2 \Bigl\{2\xi \left[(\delta^{\mu}_{\nu}R_{\sigma\tau}
   -{R^{\mu}}_{\sigma\nu\tau})\phi^{;\sigma}\phi^{;\tau}
   -R^\mu_\sigma \phi_{;\nu} \phi^{;\sigma}
  -R^\sigma_\nu \phi^{;\mu} \phi_{;\sigma}\right]
 \nonumber \\ && \hspace{1cm}
   + \xi \left[G^{\mu}_{\nu}\phi^{;\sigma} \phi_{;\sigma}
   + R \phi^{;\mu}\phi_{;\nu}\right]
   +\left(\Box \xi \,\phi^{;\sigma}\phi_{;\sigma}
   -\xi^{;\sigma\tau}\phi_{;\sigma}\phi_{;\tau}\right)
   \delta^{\mu}_{\nu}
 \nonumber \\ &&\hspace{1cm}
   +\xi^{;\mu\sigma}\phi_{;\nu}\phi_{;\sigma}
   +\xi_{;\nu\sigma}\phi^{;\mu}\phi^{;\sigma}
   -{\xi^{;\mu}}_{\nu}\phi^{;\sigma}\phi_{;\sigma}
   -\Box \xi \,\phi^{;\mu}\phi_{;\nu}
 \nonumber \\ &&\hspace{1cm}
   +\Box\phi(\xi^{;\mu}\phi_{;\nu}+\xi_{;\nu}\phi^{;\mu})
   +2\delta^{\mu}_{\nu}\left(\xi^{;\sigma}\phi^{;\tau}\phi_{\;\sigma\tau}
   -\Box\phi\,\xi^{;\sigma}\phi_{;\sigma}\right)
   +2\xi^{;\sigma}\phi_{;\sigma}{\phi^{;\mu}}_{\nu}
 \nonumber \\ &&\hspace{1cm}
   -\left[\xi_{;\sigma}\phi^{;\mu\sigma}\phi_{;\nu}
   +\xi^{;\sigma}\phi^{;\mu}\phi_{;\nu\sigma}
   +\xi^{;\mu}\phi_{;\nu\sigma}\phi^{;\sigma}
   +\xi_{;\nu}\phi^{;\mu\sigma}\phi_{;\sigma}\right]
 \nonumber \\ &&\hspace{1cm}
   +\xi\left[ 2\Box \phi\;{\phi^{;\mu}}_{\nu}
   -\delta^{\mu}_{\nu}(\Box \phi)^2
   +\delta^{\mu}_{\nu}\phi^{;\sigma\tau}\phi_{;\sigma\tau}
   -2 \phi^{;\mu\sigma}\phi_{;\nu\sigma}\right]\Bigr\}
 \nonumber \\ &&
   + c_{3} \Bigl\{2\xi \left[\phi^{;\mu\sigma}\phi_{;\nu}
   \phi_{;\sigma}+\phi^{;\mu} {\phi_{;\nu}}^{\sigma}
   \phi_{;\sigma}-\delta^{\mu}_{\nu}\phi^{;\sigma\tau}
   \phi_{;\sigma}\phi_{;\tau}-\Box\phi\,\phi^{;\mu}
   \phi_{;\nu} \right]
 \nonumber \\ && \hspace{1cm}
   +\phi^{;\sigma} \phi_{;\sigma} \left[\xi^{;\mu} \phi_{;\nu}
   +\xi_{;\nu} \phi^{;\mu} - \delta^{\mu}_{\nu} \xi^{;\tau}
   \phi_{;\tau} \right] \Bigr\}
 \nonumber \\ &&
   + c_{4} \xi \phi^{;\sigma} \phi_{;\sigma}
   \left[\delta^\mu_\nu
   \phi^{;\tau} \phi_{;\tau}-4\phi^{;\mu}\phi_{;\nu}\right]~, \\
 && \nonumber \\
 \aleph^\mu_\nu &\equiv &
   {1\over2}\delta^\mu_\nu R^{2}_{GB}
   + 4 R^\mu_{\;\;\sigma\nu\tau} R^{\sigma\tau}
   - 2 R^{\mu}_{\;\;\rho\sigma\tau}R_\nu^{\;\;\rho\sigma\tau}
   + 4 R^{\mu\sigma}R_{\nu\sigma}
   - 2 R R^\mu_\nu~, \label{e:GFE_corr} \\
 && \nonumber \\
 {\Delta}^{(c)}_{\phi} &\equiv&
  c_1 \xi_{,\phi} R^2_{GB} + c_2 G^{\mu\nu} \left[\xi_{,\phi}
  \phi_{;\mu} \phi_{;\nu} - 2 \xi_{;\mu}\phi_{;\nu}
  -2 \xi \phi_{;\mu\nu} \right]
  \nonumber \\ & &
  + c_3 \left[ \left( \xi_{,\phi} \Box\phi
  + \Box \xi\right)\phi^{;\mu}\phi_{;\mu}
  - 2 \xi^{;\mu} \phi_{;\mu} \Box \phi+ 4 \xi^{;\mu} \phi^{;\nu}
  \phi_{;\mu\nu}\right]
  \nonumber \\ & &
  +2 c_3 \xi \left[ R^{\mu\nu}\phi_{;\mu}\phi_{;\nu}
  + \phi^{;\mu\nu}\phi_{;\mu\nu}
  -(\Box \phi)^2 \right]
  \nonumber \\ &&
  + c_4 \left[ \xi_{,\phi} (\phi^{;\mu} \phi_{;\mu})^2
  -4 \phi^{;\mu} \phi_{;\mu} \left(\xi^{;\nu} \phi_{;\nu}
  + \xi \,\Box \phi\right)
  -8 \xi \phi^{;\mu} \phi^{;\nu}\phi_{;\mu\nu}  \right]~.
  \label{e:EOM_corr}
\end{eqnarray}

Eqs.~(\ref{e:G-def-Einstein-eff})--(\ref{e:EOM_corr}) form a complete
set of covariant generalised Einstein equations for a background
evolution including higher-order corrections in the metric and the
dilaton field, hence extending their domain of validity to
highly-curved regimes. For string solutions of physical interest, we
wish to restrict to four physical spacetime dimensions, which we
assume to be described by a FL metric with infinitesimal line
element $ds^2 = a^2(\eta) \left( -d\eta^2 + d\vec{x}^2 \right)$, where
$\eta$ denotes the conformal time and $~'\equiv d/d\eta$.  Inserting
this line element in the above covariant equations yields a closed
system of dynamical equations for the background evolution
(cf.~Appendix~\ref{a:background}).  Since the quantity
$\aleph^\mu_\nu$ encompasses a quadratic expression which vanishes in
four spacetime dimensions on account of the algebraic identities
satisfied by the Riemann tensor, we shall neglect it hereafter.

\begin{figure}[t]
 \begin{center}
 \includegraphics[width=0.40\linewidth]{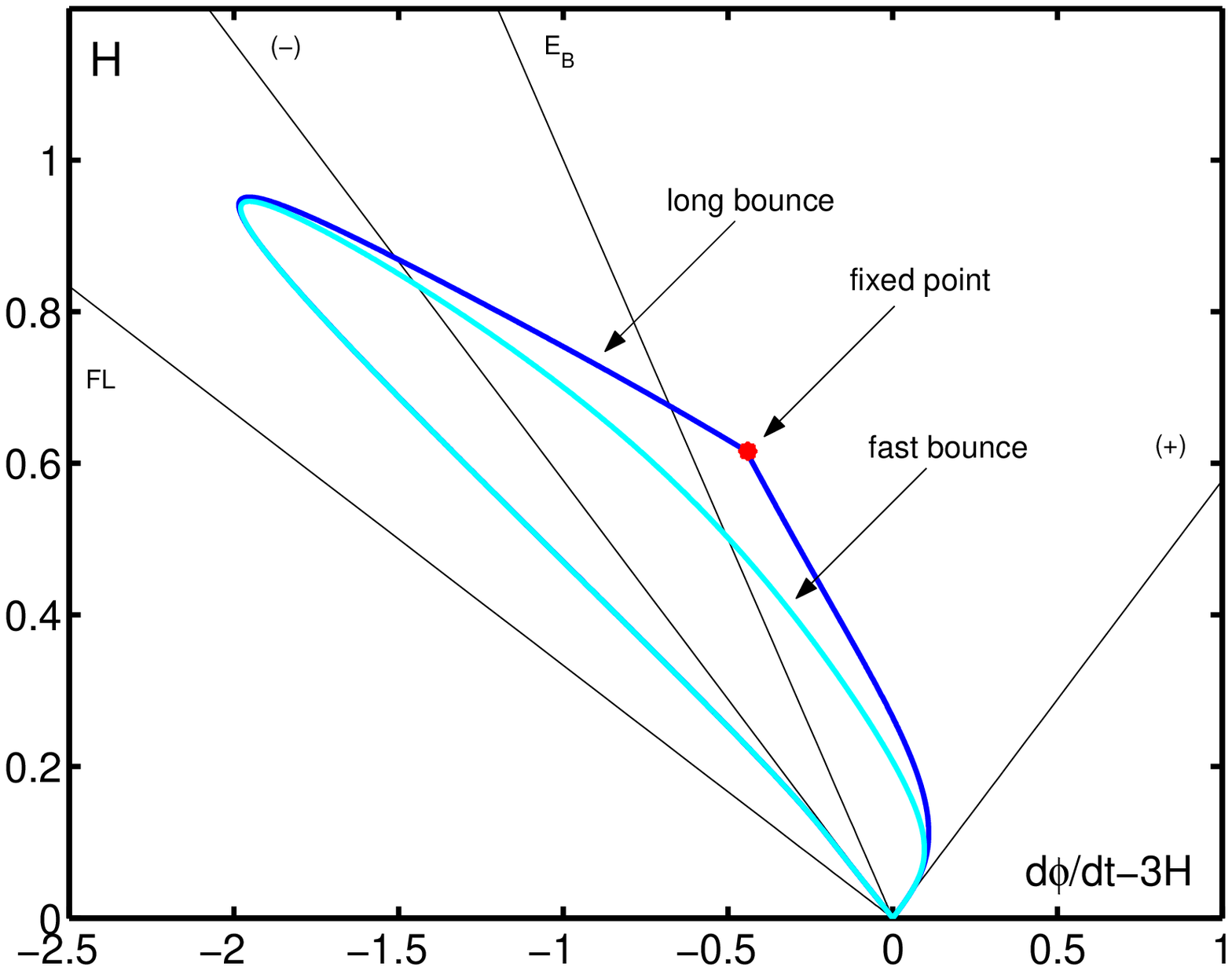}
 \quad
 \includegraphics[width=0.40\linewidth]{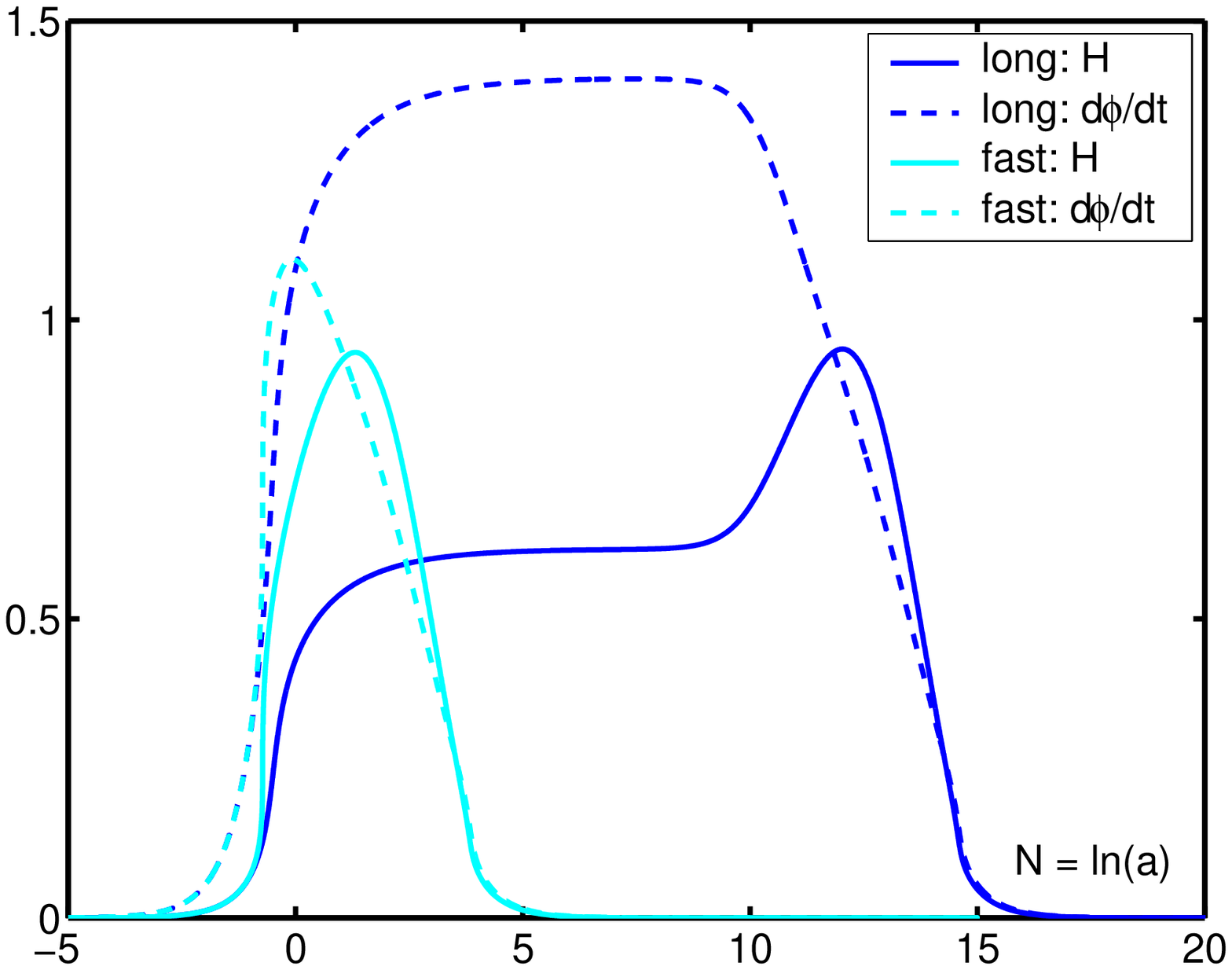}
\caption{We illustrate here two non-singular cosmological
evolutions relying on the inclusion of higher-order curvature
corrections. The figure on the left illustrates the flow diagram
$(\dot{\overline{\phi}},H)$, where the (cosmic) time coordinate is
just a parameter along the flow lines. The x-axis corresponds to
$\dot{\overline{\phi}}=\dot\phi-3H$ and the y-axis to
$H=\dot{a}/a$, the Hubble expansion rate in cosmic time in the
string frame. The $(+)$/$(-)$ branches correspond to the
accelerated/decelerated phases of the pre-big bang scenario
without potential. The line ${\rm E}_{{\rm B}}$ represents a null
Hubble parameter in the Einstein frame, $H_e = 0$. A typical
non-singular evolution thus emerges from a weak-coupling and
low-curvature state, located at the origin of the flow diagram.
Then follows a dilaton-driven phase with increasing coupling and
curvature along the $(+)$ branch and runs into a perturbative
singularity, unless higher-order corrections force the background
to undergo a branch change, $\dot{\overline\phi} \geqslant 0 \to
\dot{\overline\phi} \leqslant 0$, hence regulating the background
evolution. Crossing over the Einstein bounce ensures that the
Hubble parameter in the Einstein frame is henceforth positive. The
figure on the right shows the corresponding evolution of $H$ and
$\dot\phi$ as a function of the number of e-folds, $N=\ln(a)$.
Both regular evolutions have been obtained for $c_1=-1$,
$c_2=c_3=0$, $c_4=1$, with a loop control function
$\xi(\phi)=e^{-\phi}+ \xi_A + \xi_B e^{\phi}$. For the fast
bounce, we set $\xi_A=1$ and $\xi_B=-2.5\cdot 10^{-3}$. The
initial conditions for the simulations have been set with respect
to the lowest-order analytical solutions at $\eta = -10^6$. A long
high curvature regime is easily obtained by trapping the
background evolution in the $\alpha'$-asymptotic fixed point
(denoted by a $\bullet$), featuring a constant Hubble rate and
linearly growing dilaton in cosmic time. This can be achieved by
reducing the constant parameters $\xi_A$ and $\xi_B$, or by
shifting the initial value of the dilaton, $\phi \to \phi-\phi_0$
with, e.g., $\phi_0=25$. Eventually the string coupling $g_s =
e^{\phi/2}$ grows strong enough to allow for a violation of the
null energy condition $\rho+p\geqslant 0$ and enables a successful
exit to the decelerated post-big bang regime.}
\label{f:background}
\end{center}
\end{figure}

As an application of these equations, we may consider in
particular the pre-big bang scenario of string cosmology
\cite{Veneziano:1991ek,Gasperini:1993em} which requires ${\cal
F}=e^{-\phi} =-\omega(\phi)$ and ${\cal V}(\phi)=0$ when
considered at tree-level in the string frame. In string theory,
the precise form of the higher-order corrections can be fixed by
requiring that our effective action Eq.~(\ref{e:general-action})
reproduces the string theory $S$-matrix elements. At the
next-to-leading order in the $\alpha'$ expansion, this only
constrains the coefficient of $R_{\mu\nu\lambda\rho}^2$ with the
result that the pre-factor for the Gauss-Bonnet term has to be
$c_1=1$. But the lagrangian can still be shifted by field
re-definitions which preserve the on-shell amplitudes, leaving the
three remaining coefficients satisfying
\begin{equation}
 \xi(\phi) = e^{-\phi}~, \qquad c_{1}=1~, \qquad
 c_3=-\frac{1}{2}\left[c_2+2\left(c_1+c_4\right)\right]~.
\end{equation}
The parameter $\lambda$ allows us to move between different string
theories~: $\lambda = 1/4,~1/8$ for the bosonic and heterotic string
respectively, whereas for type II superstrings $\lambda =0$ and
corrections start at higher order \cite{Gross:1986iv,Metsaev:1987zx}.
Here we shall use $\lambda =1/4$ to agree with previous
studies~\cite{Gasperini:1997fu}. The natural setting $c_2=c_3=0$ leads
to the well-known form which has given rise to most of the studies on
corrections to the low-energy picture.
In~\cite{Gasperini:1997fu,Brustein:1998cv}, the authors demonstrated
that this set of minimal tree-level $\alpha'$ corrections regularises
the singular behaviour of the low-energy pre-big bang scenario.
Emerging from the asymptotic past vacuum along the low-energy exact
pre-big bang solution, the $(+)$ branch $\phi' = \left(3+\sqrt{3}
\right) {\cal H}$ with increasing coupling and curvature, the
$\alpha'$ corrections drive the evolution to a fixed point of bounded
curvature with a linearly growing dilaton in cosmic time (the
$\bullet$ in the left panel of Fig.~\ref{f:background}). This suggests
that quantum loop corrections --- known to allow a violation of the
null energy condition, $p+\rho \geqslant 0$ --- would permit the
crossing of the Einstein bounce ${\cal H}_e = {\cal H}
-\phi'/2$~\cite{Brustein:1997ny} (the subscript -e- denoting a
quantity evaluated in the Einstein frame) and a graceful exit to a FL
decelerated expansion in the post-big bang era, represented by the
$(-)$ branch with $\phi'=\left(3-\sqrt{3}\right){\cal H}$. Little is
known about the exact form of quantum loop corrections and we take the
freedom to parameterise them with an expansion in the string coupling
$g^2_s = e^{\phi}$ of the form $\xi(\phi) = e^{-\phi} + \xi_A + \xi_B
e^{\phi}$ with constant free parameters $\xi_A$ and $\xi_B$. An
example of successful exit is shown in Fig.~\ref{f:background}. In
general, the combination of tree-level $\alpha'$ and quantum loop
corrections does not lead to a constant value for the dilaton in a
finite amount of time. But this can be fixed by introducing a suitable
potential or invoking particle production
\cite{Brustein:1998cv,Cartier:1999vk}.

Given that there exists a class of non-singular cosmologies based
on these higher-order corrections, it is then natural to
investigate the effect of the correction terms on the evolution of
primordial scalar fluctuations.

\section{Linear perturbations}
\label{s:G-perturbations}

\subsection{The curvature perturbation ${\cal R}$}
\label{ss:curvature-pert}

To study perturbations, we include small deviations of the
background metric, $g_{\mu\nu}= \bar{g}_{\mu\nu} +\delta
g_{\mu\nu}$, and for the scalar field, $\phi$. We know
from~\cite{Cartier:2001is} that an homogeneous second-order
differential equation for the curvature perturbation on uniform
energy density hypersurfaces, ${\mathcal R}$, can be derived from
the general action Eq.~(\ref{e:general-action}) including
higher-order corrections Eq.~(\ref{e:action-corr-gen}). To linear
order, the decomposition in Fourier modes implies that each
comoving wavenumber $k$ evolves independently from other comoving
modes according to
\begin{equation}
 {\cal R}'' + 2\frac{z'}{z}{\cal R}' +s k^2 {\cal R}= 0~,
 \label{e:G-wave-R}
\end{equation}
where we used the eigenstates of the Laplace-Beltrami operator,
$\Delta{\cal R}=- k^2{\cal R}$. Here, we introduced two functions of
time, $z = a\sqrt{Q}$ and $s$. For a minimally coupled massless field
$\phi$ and neglecting the contribution of the curvature corrections,
these parameters reduce to $Q \to (\phi'/{\cal H})^2$ and $s \to +1$.
However, in the more general situation where the coupling remains
unspecified and we allow for higher-order terms of the form of
Eq.~(\ref{e:action-corr-gen}), we have (see~\cite{Cartier:2001is} for
an explicit derivation)
\begin{eqnarray}
 Q &\equiv&
       \frac{\omega\phi^{\prime 2}
       +3\frac{\left({\cal F}'+aQ_{a}\right)^2}{2{\cal F}+Q_{b}}+a^2Q_{c}}
       {\left[{\cal H}+\frac{{\cal F}'+a Q_{a}}{2{\cal F}+Q_{b}}\right]^{2}}
  ~=~ \frac{\phi'E_{2}}{a^2E_{1}^2}~\Omega~, \qquad
  \Omega(\eta) ~\equiv~ E_2 E_5 + 3 E_1 E_4~,
       \label{e:G-Q-def-R} \\
   && \nonumber \\
 s &\equiv&
       1 + \frac{a^2Q_{d}+a\frac{{\cal F}'+aQ_{a}}{2{\cal F}+Q_{b}}Q_{e}
       +\left[\frac{{\cal F}'+aQ_{a}}{2{\cal F}+Q_{b}}\right]^2Q_{f}}
       {\omega\phi^{\prime 2}+3\frac{\left({\cal F}'+aQ_{a}\right)^2}
       {2{\cal F}+Q_{b}}+a^2Q_{c}}~.
       \label{e:G-s-def-R}
\end{eqnarray}
To be concise, the $Q_i$'s and $E_i$'s are complicated functions
of the background time-dependent quantities, the $Q_i$'s
representing the $\alpha'$ modifications of the tree-level
equation. Their explicit form is given in the Appendix.
Introducing the canonical variable $v \equiv z{\cal
R}$~\cite{Mukhanov:1992me}, the linearised wave equation
(\ref{e:G-wave-R}) reduces to
\begin{equation}
   v'' + \left[sk^2 - U(\eta) \right] v = 0~,\qquad
   U(\eta) = \frac{z''}{z} ~.
   \label{e:G-wave-canonical-v}
\end{equation}
Through the external potential $U(\eta)$, the ``pump'' field $z$
is responsible for the parametric amplification
\cite{Grishchuk:1990bj,Grishchuk:1992tw} of the metric
fluctuations. In the case where the background evolution undergoes
a period of pole-like inflation (e.g., derived from induced
gravity \cite{Pollock:1989vn}, scalar-tensor
gravity~\cite{Levin:1994wr}, the (modified) pre-big-bang scenario
\cite{Veneziano:1991ek,Gasperini:1993em,Durrer:2002jn} or the
ekpyrotic model~\cite{Khoury:2001wf,Khoury:2001bz,Khoury:2001zk}),
the pump field reduces to a power-law, $z\propto a_e =|\eta|^q$,
hence $U(\eta) =q(q-1)|\eta|^{-2}$. Then
Eq.~(\ref{e:G-wave-canonical-v}) corresponds for $s=1$ to a Bessel
equation and its general solution can be expressed as a
superposition of Hankel functions of the first and second kind
which, in terms of our original variable, reads
\cite{Cartier:2001is}
\begin{equation}
 {\cal R}=\frac{\sqrt{\pi|\eta|}}{2z}\sum_{i=1}^2 d_i(k)H_\nu^{(i)}(k |\eta|)~,
 \qquad \nu\equiv\frac{1}{2}\left|1-2 q\right|~.
  \label{e:G-Bessel-sol-R}
\end{equation}
This solution satisfies the usual unitary condition between the
coefficients $|d_2|^2 - |d_1|^2 = 1$, while choosing a pure
positive frequency state in the asymptotic flat space time for
$\eta \rightarrow -\infty$ requires $d_2 = 1$ and $d_1 = 0$.

In the large scale limit $k/|{\mathcal H}|\to 0$,
Eq.~(\ref{e:G-Bessel-sol-R}) reduces to ${\mathcal R} \simeq A(k)
+B(k) |\eta|^{1-2q}$, where the coefficients $A$ and $B$ are fixed
by the exact solution~(\ref{e:G-Bessel-sol-R}).  Defining the
initial power spectrum by ${\mathcal P}_x \equiv |x|^2k^3 \propto
k^{n_x-1}$, the spectral index carried by each mode before the
bounce is thus $n_A = 3+2q$ and $n_B = 5-2q$. For the pre-big bang
scenario with vanishing potential, we have $q=1/2$ and
$n_A=n_B=4$. However, adiabatic density fluctuations can also be
described by the gauge-invariant Bardeen potential in the
longitudinal gauge, $\Phi_e = \Phi-\frac{1}{2}\delta\phi_\chi$,
which evolves according to $\Phi_e'' + 6{\mathcal H}\Phi'_e +
k^2\Phi_e=0$ at tree-level in the Einstein frame. In the large
scale limit, $\Phi_e \simeq C(k)|\eta|^{-(1+2q)}+D(k)$ and
consistent normalisation yields $n_C=1-2q =0$ and $n_D=3+2q=4$ for
the pre-big bang scenario. Comparing the amplitudes of the modes
$C$ and $D$, we clearly have $\Phi_{e,C}\gg \Phi_{e,D}$ right
before the bounce, $|\eta|\to 0$ (see~\cite{Cartier:2003jz} for
the details). Naively, we therefore expect that the dominant
spectral index after the bounce remains $n_C$.  However, since in
the radiation dominated era $\Phi_e$ and ${\mathcal R}$ differ
only by a $k$-independent constant, general relativity requires
them to have the same spectral distribution. Hence the claims that
the spectral distribution of ${\mathcal R}$ is unaltered during a
bounce and that the dominant pre-bounce spectral index yields the
late time post-bounce spectral distribution for $\Phi_e$ are
contradictory.

One possibility to discriminate among these incompatible results
is to consider a regular background evolution. As discussed
in~\cite{Cartier:2001is}, the
corrections~(\ref{e:action-corr-gen}) genuinely alter the
time-dependence of the variables $z=a\sqrt{Q}$ and $s$ during the
high-curvature regime; the source $Q$ of the effective potential
for the perturbation may no longer have a monotonic growth and may
even decrease at the onset or during a high-curvature regime.
Apparently, there is also no restriction on the sign of the
frequency shift occurring when the curvature corrections dominate
the background dynamics: $s$ may become negative or infinite
depending on its particular form. Previous applications of
Eqs.~(\ref{e:G-wave-R})--(\ref{e:G-s-def-R}) to pre-big bang-like
scenarios~\cite{Cartier:2001is,Tsujikawa:2002qc} have however
missed out the singular behaviour of $Q$ during a graceful exit.
Indeed the components of the $Q$-denominator $E_1= a^2\left({\cal
F}'+ 2{\cal F}{\cal H}+aQ_{g}\right)$ have to cancel out each
other at some particular time, since the asymptotic analytical
branches of the pre-big bang scenario (${\mathcal F}=e^{-\phi}$ in
the string frame) yields opposite signs for $E_1$~:
\begin{eqnarray}
 -\infty < \eta < -\eta_s~:~~\phi'=(3+\sqrt{3}){\mathcal H}
 &~\rightarrow~& E_1 = -(\sqrt{3}+1)a^2{\mathcal H}e^{-\phi} < 0~,
 \label{e:sol-preBB}\\
 \eta_s < \eta < +\infty~:~~\phi'=(3-\sqrt{3}){\mathcal H}
 &~\rightarrow~& E_1 = (\sqrt{3}-1)a^2{\mathcal H}e^{-\phi} > 0~,
 \label{e:sol-postBB}
\end{eqnarray}
where $\eta_s$ represents the time at which higher-order corrections
dominate the background dynamics. Furthermore, this is the case for
any FL spatially-flat bouncing universe which is described by the
action Eq.~(\ref{e:general-action}), as $E_1 \propto {\cal H}-\phi'/2
= {\cal H}_e$ has to change sign during the transition between a
collapsing phase and an expanding phase. Therefore the pump field $z$
and the ratio $z'/z$ are singular, which invalidates
the use of Eq.~(\ref{e:G-wave-R}) in the context of bouncing universes
of the type considered here.

We now proceed to investigate yet another dynamical equation for
the perturbation ${\cal R}$, for we are interested in obtaining
the spectrum long after the transition from a regular equation.

\subsection{Coupled systems $(\Phi,\delta\phi_\chi)$ and $(\Phi,{\cal R})$}
\label{ss:coupled-systems}

The approach we consider henceforth relies on the gauge-invariant
Bardeen potentials, $\Phi$ and $\Psi$, and the perturbation in the
scalar field, $\delta\phi_\chi$. Here we shall present our main
results, and leave the definition of these variables and an
explicit derivation of the terms involved to
Sec.~\ref{a:perturbation}. At tree-level, it is well
known~\cite{Mukhanov:1992tc} that these variables satisfy
homogeneous second-order differential equations at all times. When
higher-order corrections are included, however, the decoupled wave
equation of each of these variables is plagued by a similar
singular behaviour during the transition regime as the one of
${\mathcal R}= -\Phi + \frac{{\mathcal H}}{\phi'}\delta\phi_\chi$.
To see this, we recall that the covariant equations
(\ref{e:G-def-Einstein-eff})--(\ref{e:G-EOM-phi}) yield five
redundant evolution equations in terms of the gauge-invariant
perturbed variables $\Phi$, $\Psi$ and $\delta\phi_\chi$. They are
the four components of the Einstein equation, $\left( ^0_0
\right)$, $\left( ^0_j \right)$, $\left( ^i_i \right)$ and the
trace-free part $(~)_{tf}=\left( ^i_j \right)-\frac{1}{3}\left(
^k_k \right)\delta^i_j$. Those are supplemented by the perturbed
part of the scalar field equation $\left(\phi\right)$. The $\left(
^0_0 \right)$ and $\left( ^0_j \right)$ equations are first order
in derivatives of the perturbed variables, while the $\left( ^i_i
\right)$ and $\left(\phi\right)$ are of second order. Of prime
interest is the tracefree part which implies a linear
background-dependent relation among the perturbation variables
\begin{equation}
 \Psi = c_\Phi \Phi +c_{\delta\phi_\chi} \delta\phi_\chi~,
 \label{e:G-Psi-higher-order}
\end{equation}
where the coefficients $c_\Phi$ and $c_{\delta\phi_\chi}$ are
functions of time (see Appendix),
\begin{equation}
 c_\Phi = -E_{3}E_{2}^{-1}~, \quad\mbox{and}\quad
 c_{\delta\phi_\chi}=-E_{6}E_{2}^{-1}~.
 \label{e:def-cPhi}
\end{equation}
In the limit $\alpha'\to0$, we recover $\Psi + \Phi =
\delta\phi_\chi$ in the string frame, which corresponds to the
usual relation $\Psi_e+\Phi_e=0$ when expressed in the Einstein
frame. The relation among these variables is modified in a
non-trivial way when the background dynamics is dominated by the
curvature corrections, but the alteration does not exhibit any
dependence with respect to the comoving wavenumber $k$. For a
regularised background evolution with $E_2 \neq 0$ at all times
(this is the case for a wide range of models we have tested),
Eq.~(\ref{e:G-Psi-higher-order}) enables us to replace the
variable $\Psi=\Psi(\Phi,\delta\phi_\chi)$ in the perturbation
equations. We can then manipulate the remaining equations to
extract a set of decoupled dynamical equations for the
perturbations $\Phi$ and $\delta\phi_\chi$. However, these
equations are plagued by a singular behaviour during the
transition regime of the kind we previously discussed for ${\cal
R}$. To demonstrate that we cannot obtain \textit{regular}
decoupled second-order wave equations for each of these variables,
we focus on the $\left( ^0_0 \right)$ and $\left( ^0_j \right)$
equations, which completely specify the dynamical evolution of the
perturbations. These first-order equations can be written as
${\cal A}X' + {\cal B}X = 0$ where $X\equiv
(\Phi,\delta\phi_{\chi})^T$, and ${\cal A}(\eta)$ and ${\cal
B}(\eta,k)$ are $2\times 2$ matrices whose coefficients are
complicated combinations of the background quantities. Provided
that the matrix ${\cal A}$ is invertible, i.e., $\det({\cal A}) =
-2\Omega \neq 0$ at all times, we can extract an homogeneous
system of linear differential equations $X'={\cal C}X$ with
time-dependent coefficients ${\cal C}(\eta,k)=-{\cal A}^{-1}{\cal
B}$~:
\begin{equation}
 \pmatrix{\Phi \cr \delta\phi_{\chi}}' =
 \pmatrix{{\cal C}_{11} & {\cal C}_{12} \cr  {\cal C}_{21} & {\cal C}_{22}}
 \pmatrix{\Phi\cr \delta\phi_{\chi}}~.
 \label{e:dyn-sys-Phi-dp}
\end{equation}
This homogeneous linear system has a set of $n=2$ linearly
independent solutions of the form
\begin{equation}
 X(\eta,k) = T \exp{\left(\int^\eta {\cal C}(\tilde{\eta},k)
 d\tilde{\eta}\right)}X(\eta_0,k)~,
 \label{e:sol-C}
\end{equation}
where $T$ is a time-ordering operator. If ${\cal C}_{ij}(\eta)$ are
continuous functions and initial conditions are prescribed, the system
has a unique solution given by Eq.~(\ref{e:sol-C}).  Although every
differential equation of order $n$ can be rewritten as a first-order
system of $n$ equations, the system viewpoint is more general as we
shall recall now. In principle, the system
Eq.~(\ref{e:dyn-sys-Phi-dp}) can be reduced to a second-order
differential equation by elimination of one of the variable, e.g.,
$\delta\phi_{\chi}$. Indeed it is straightforward to obtain the
homogeneous differential equation
\begin{equation}
 \Phi''-\left({\cal C}_{11}+{\cal C}_{22}
 +\frac{{\cal C}_{12}'}{{\cal C}_{12}}\right)\Phi'
 +\left[{\cal C}_{11}{\cal C}_{22}-{\cal C}_{12}{\cal C}_{21}
 +\frac{1}{{\cal C}_{12}}\left({\cal C}_{11}{\cal C}_{12}'
 -{\cal C}_{11}'{\cal C}_{12} \right) \right]\Phi = 0~.
 \label{e:Phi-second-order}
\end{equation}
But if ${\cal C}_{11}$ or ${\cal C}_{12}$ are not differentiable,
or if ${\cal C}_{12}=0$ at some particular time $\eta_{*}$, the
reduction to Eq.~(\ref{e:Phi-second-order}) will not be possible
(leaving aside the trivial case for which  ${\cal C}_{12}=0$ at all
time).

For all background evolutions of the type discussed in
Sec.~\ref{s:action-GenGravity} we have tested, we find that $E_{2}
\neq 0$ and $\Omega \neq 0$ during the whole evolution, while ${\cal
 C}_{12}$ and ${\cal C}_{21}$ change sign. Hence
homogeneous second-order equations valid at all times for either
$\Phi$ or $\delta\phi_\chi$ cannot be derived in our particular class
of cosmologies, and we shall use Eq.~(\ref{e:dyn-sys-Phi-dp}) to
determine the spectral distribution long after the high-curvature
regime. Explicitly, the decomposition ${\cal C}_{ij}(\eta,k) =
\hat{\cal C}_{ij}(\eta) + k^2 \breve{\cal C}_{ij}(\eta)$ yields
\begin{equation}
 \hat{\cal C}(\eta) =
 \pmatrix{{\cal H}c_{\Phi} & {\cal H}c_{\delta\phi}-\Xi \cr
 \phi'c_{\Phi} & \phi'c_{\delta\phi}+\Pi}~, \qquad
 \breve{\cal C}(\eta) = \Omega^{-1}
 \pmatrix{-E_{2}E_{4} & -E_{4}^2 \cr E_{2}^2 & E_{2}E_{4}}~,
 \label{e:matrix-C-C}
\end{equation}
where the coefficients $c_{\Phi}(\eta)$ and $c_{\delta\phi}(\eta)$
are given in Eq.~(\ref{e:def-cPhi}), $\Omega(\eta)$ is given by
Eq.~(\ref{e:G-Q-def-R}) and we define
\begin{eqnarray}
 \Xi(\eta) &\equiv& \left(E_4 E_{10}+2E_5E_{11}\right)/(2\Omega)~,\\
 \Pi(\eta) &\equiv& -\left[\Xi+\left({\cal H}/\phi'\right)'\right]
 \phi'/{\cal H}~.
\end{eqnarray}

Hence, Eqs.~(\ref{e:dyn-sys-Phi-dp}) and~(\ref{e:matrix-C-C})
provide us with the necessary tools to determine the spectral
distribution of $\Phi$ and $\delta\phi_\chi$ long after the
high-curvature regime. If numerical simulations were close to
infinite precision, we could then deduce the evolution of ${\cal
R}= -\Phi + \frac{{\cal H}}{\phi'} \delta\phi_\chi$ and its late
time power spectrum. This in turn would enable us to discriminate
among the incompatible claims resulting from different matching
procedures used in the tree-level analysis. Unfortunately, lack of
numerical precision inevitably spoils the spectral distribution of
${\cal R}$. In a string cosmology background, we recall that
metric and field perturbations are normalised with respect to the
maximal amplified frequency $k_{max}\equiv \textrm{Max}({\cal
H})$, for which the perturbation variables have similar
amplitudes, $|\delta_\Phi| \approx |\delta_{\delta\phi_\chi}|
\approx |\delta_{\cal R}|$. Therefore, as long as $\Phi$ and
$\delta\phi_\chi$ carry red spectra, we will have $|\delta_\Phi|
\approx |\delta_{\delta\phi_\chi}| \approx (k_{max}/k)^2
|\delta_{\cal R}|$ since we expect ${\cal R}$ to have a steep blue
$k^3$ spectrum. But for sufficiently small wavenumbers, $k\ll
k_{max}$, the spectrum of ${\cal R}$ obtained from ${\cal R}=
-\Phi + \frac{{\cal H}}{\phi'} \delta\phi_\chi$ will be dominated
by the imprecision in numerically evolving $\Phi$ and
$\delta\phi_\chi$~; the deduced spectrum of ${\cal R}$ would thus
be red, i.e., a measure of the inaccuracy of the numerical
simulations. To avoid such contamination and erroneous conclusion,
we may instead consider the variable $Y \equiv (\Phi,{\cal R})$,
whose dynamics also satisfies an homogeneous system of linear
differential equations of first order, $Y'={\cal D}(\eta,k)Y$~:
\begin{eqnarray}
 \pmatrix{\Phi \cr {\cal R}}' =
 \pmatrix{
 {\cal H}c_{\Phi}+\phi'c_{\delta\phi}-\frac{\phi'}{{\cal H}}\Xi&
 \phi'c_{\delta\phi}-\frac{\phi'}{{\cal H}}\Xi\cr 0&0}
 \pmatrix{\Phi\cr {\cal R}}
 + \frac{k^2}{\Omega{\cal H}\phi'}
 \pmatrix{ -E_{1}E_{4}\phi' & -E_{4}^2\phi^{\prime 2} \cr
 E_{1}^2 & E_{1}E_{4}\phi'} \pmatrix{\Phi \cr {\cal R}} ~,
 \label{e:dyn-sys-Phi-R}
\end{eqnarray}
where we have decomposed again the matrix elements according to ${\cal
  D}_{ij}(\eta,k) =\hat{\cal D}_{ij}(\eta) + k^2 \breve{\cal
  D}_{ij}(\eta)$. Eq.~(\ref{e:dyn-sys-Phi-R}) is the main result of
this analysis~: It determines the evolution of the curvature
perturbation ${\cal R}$ on super-Hubble scales. Indeed, the background
equations, including those higher-order corrections needed to
regularise the background evolution, imply at all times the exact
cancellation of some coefficients, such that $\hat{\cal D}_{21}\equiv
0 \equiv\hat{\cal D}_{22}$. Hence ${\cal R}$ remains nearly constant
on super-Hubble scales, its evolution entering only as a $(k/{\cal
  H})^2$ correction, while Bardeen's potential may evolve drastically
on super-Hubble scales. This clearly suggests that, if the
perturbation variables ${\cal R}$ and $\Phi$ are to yield the same
spectral index long after the transition, the pre-bounce growing mode
of the Bardeen potential has to be fully converted into its decaying
mode during the high-curvature regime. This is confirmed by noting
that the leading source term of $\Phi'$ on super-Hubble scales, i.e.,
the coefficient $\hat{\cal D}_{11}$ goes from positive to negative
during the high-curvature regime. The growth of the Bardeen potential
is thus turned into a rapid decay during the high-curvature regime.

The complexity of the matrix elements of
Eq.~(\ref{e:dyn-sys-Phi-R}) does not enable us to solve this
system of differential equations exactly, and we have to resort to
numerical integration of the perturbation equations to confirm our
prediction about the post-bounce spectral distributions. Initial
conditions in the asymptotic past can be set according to
\begin{equation}
 \pmatrix{\Phi \cr \delta\phi_\chi} = \frac{\phi'}{E_1^2}
 \pmatrix{-E_1E_4 & k^{-2}{\cal H}\Omega \cr E_1E_2 & k^{-2}\phi'\Omega}
 \pmatrix{{\cal R} \cr {\cal R}'}~,
     \label{e:G-relation-Phi-dpx-versus-R-R'}
\end{equation}
where we may use the exact tree-level pre-big bang solution
(\ref{e:G-Bessel-sol-R}) for ${\cal R}$ and its first time derivative.
We recall that these relations cannot be used at all times since
$E_{1}$ is forced to change sign during the course of the background
evolution, and we have also shown that there is no second-order
homogeneous equation for ${\cal R}$ valid at all time. However these
relations are fully adequate to initialise the perturbations $\Phi$
and $\delta\phi_\chi$ at early times $\eta \ll -\eta_s<0$.

\begin{figure}[t]
 \begin{center}
 \includegraphics[width=0.45\linewidth]{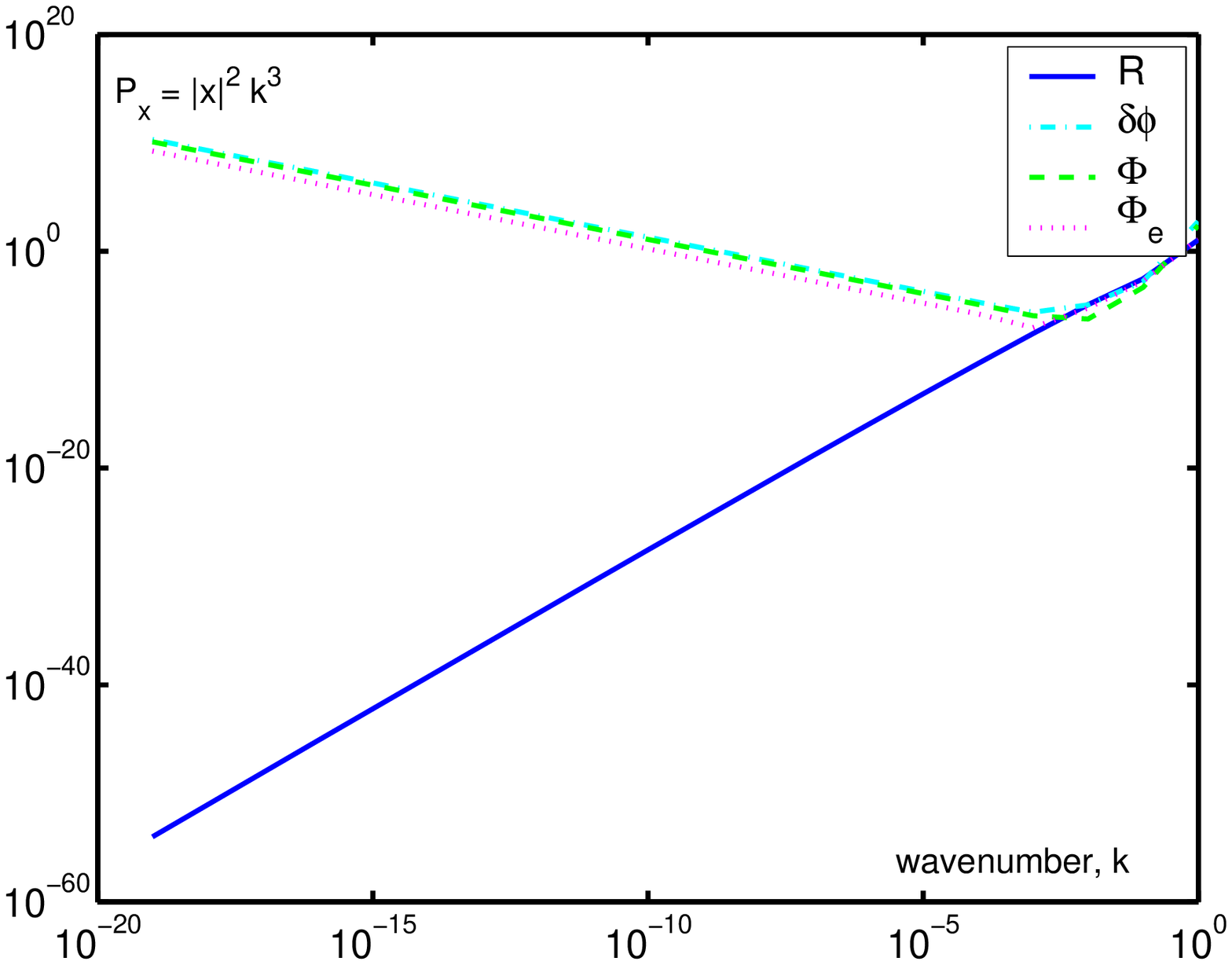}\quad
 \includegraphics[width=0.45\linewidth]{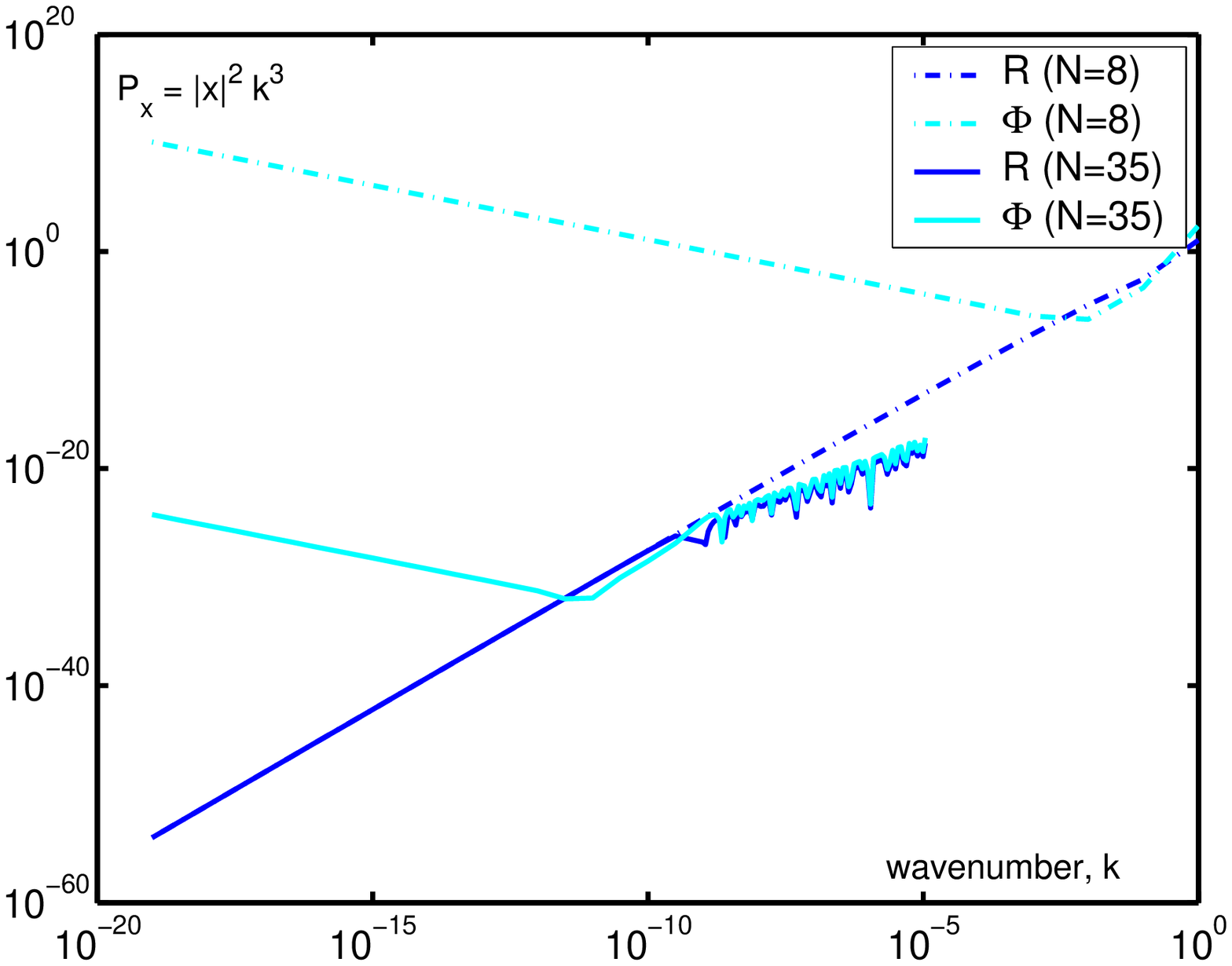}
 \caption{We illustrate here the post-bounce spectral distributions
   of adiabatic density perturbations in the non-singular cosmological
   background of Fig.~\ref{f:background}. We considered the
   high-curvature transition of short duration in order to reduce
   numerical uncertainty. The picture on the left shows ${\cal P}_{x}
   = |x|^2 k^3$ for $x\in \{ {\cal R}, \delta\phi_\chi, \Phi, \Phi_e
   \}$ evaluated shortly after the bounce, $N =\ln(a)\simeq 8$.  The
   wavenumber $k$ is given in units of $k_{max} \equiv
   \textrm{Max}({\cal H})$.  The red spectrum attached to the decaying
   mode of $\Phi$ is still dominant on nearly all scales, while ${\cal
     R}$ carries a blue spectrum. On the panel on the right we compare
   the spectral distributions ${\cal P}_{x} = |x|^2 k^3$ for $x\in \{
   {\cal R}, \Phi\}$ evaluated at two different times after the
   bounce~: $N\simeq 8$ and $N\simeq 35$. The amplitude of ${\cal
     P}_{\Phi}$ is decreasing in time until it reaches its
   sub-dominant constant mode, which carries a blue spectrum. Hence,
   both perturbation variables $\Phi$ and ${\cal R}$ lead to the same
   blue primordial power spectrum long after the transition. Finally,
   back inside the Hubble radius, the spectra start to oscillate
   and we have $\Phi = \pm \sqrt{3}{\cal R} + \textrm{cst}$.}
 \label{f:spectrum}
 \end{center}
\end{figure}

Figure~\ref{f:spectrum} illustrates the results of numerical
  integration based on the system Eq.~(\ref{e:dyn-sys-Phi-R}) for the
  pre-big bang scenario of string cosmology. We observe that the
  pre-bounce dominant mode of the Bardeen potential carrying the red
  spectrum is fully converted into the decaying post-bounce mode.
  Indeed, as shown in~\cite{Cartier:2003jz}, a kink in the spectral
  distribution arises only in the situation where the growing mode of
  the pre-bounce phase is fully converted into the decaying mode after
  the transition, and one has to wait a sufficiently long time for the
  decaying mode to decay and the final growing mode to dominate. As a
  result, the pre-bounce subdominant mode yields, long after the
  bounce, the spectral index relevant for the observed anisotropies in
  the cosmic microwave background; it corresponds to a steep blue
  spectrum with $n_\Phi=3+2q = 4$. Hence, although the perturbation
  variables $\Phi$ and ${\cal R}$ carry different spectral
  distributions before the bounce, our numerical simulations confirm
  that they have the same spectral index sufficiently long after the
  transition when the decaying mode of $\Phi$ has died away. This
  favours the argument put forward
  in~\cite{Brandenberger:2001bs,Lyth:2001pf,Hwang:2001ga}, where the
  authors have shown that this is exactly what happens if a matching
  between the pre- and post-bounce epochs is defined by a vanishing
  jump in the metric and the second fundamental form on the constant
  energy hypersurface.

\section{Conclusion}
\label{s:conclusion} Even though we have not been able to
demonstrate from first principles that the pre-bounce growing mode
of the Bardeen potential $\Phi$ (which carries a red spectrum) has
to be fully converted into the decaying mode during the
high-curvature regime, we have derived a new master equation which
enables us to describe the evolution of adiabatic density
perturbations on super-Hubble scales. This master equation takes
the form of an homogeneous system of linear differential equations
of first order, explicitly coupling the Bardeen potential $\Phi$
and the curvature perturbation ${\cal R}$. On the one hand, we
have shown that such system cannot be reduced to second order
homogeneous differential equations for the class of cosmologies
considered here, hence correcting erroneous claims in the
literature. On the other hand, this master equation clearly
indicates that the curvature perturbation is nearly constant on
super-Hubble scales, its time-dependence entering as a $(k/{\cal
H})^2$ correction. This implies that the pre-bounce growing mode
of $\Phi$ is fully converted into the decaying mode during the
transition. This is confirmed by numerical simulations which
indicate that the dominant mode of the Bardeen potential carries a
blue spectrum long after the transition. As we do not expect that
a smooth transition from the decelerating post-big bang phase to
the usual radiation-dominated FL epoch ($\phi=$ cst.)  modifies
the spectral distribution, we find that the final spectrum of
adiabatic density perturbation is strongly blue tilted for the
class of cosmologies considered here. This result is in complete
agreement with~\cite{Gasperini:2003pb} where the bounce is
triggered by a non-local potential instead of higher-order curvature
corrections.

\section*{Acknowledgments}
We are grateful to Ruth Durrer and Peter Wittwer for stimulating
and clarifying discussions. The author acknowledges financial
support from the Tomalla Foundation and the Swiss National Science
Foundation.



\providecommand{\href}[2]{#2}

\appendix

\section{}
In this appendix, we provide the necessary tools to reproduce the
results discussed in the main text. After introducing our
notation, we present the different components entering the
background equations of motion. Then we derive the gauge-invariant
expressions which contribute to the linear perturbation equations.
Finally, we introduce our matrix notation, and give all the
relevant terms in order to derive the systems of differential
equations governing the evolution of adiabatic density
perturbations.

\subsection{Notation}
\label{a:notation}
To linear order the most general perturbation in a
four-dimensional FL spacetime is given by the infinitesimal
line element
\begin{eqnarray*}
 d s^2 &=&
 - a^2 \left( 1 + 2 \tau \right) d \eta^2
 - 2 a^2 \left( \beta_{|i} + B_i \right) d \eta d x^i
 + a^2 \left [ g^{(3)}_{ij} \left( 1 + 2 \varphi \right)
 + 2 \gamma_{|ij} + 2C_{(i|j)} + 2 h_{ij} \right ] d x^i d x^j ~.
 \label{e:metric_pert}
\end{eqnarray*}
Under the gauge transformation
\begin{eqnarray}
   \eta \mapsto \eta + \zeta^0 (\eta,{\bf x}) ~,
    \label{e:gauge_trans1} \qquad
   x^i \mapsto x^i + g^{(3)\,ij}
    \, \zeta^{(s)}_{\;\;\;\;\;|j}(\eta,{\bf x})
    + \zeta^{(v)\,i}(\eta,{\bf x}) ~,
\end{eqnarray}
the induced changes in the perturbed part of the metric read
\begin{eqnarray}
 && \tau \mapsto \tau - {\cal H} \zeta^0 - \zeta^{0\,\prime} ~,
    \hspace{0,7cm}
    \varphi \mapsto \varphi - {\cal H} \zeta^0 ~,
    \hspace{0,8cm}
    \gamma \mapsto \gamma -\zeta^{(s)} ~,
    \hspace{0,7cm}
    h_{ij} \mapsto h_{ij} ~,
    \nonumber  \\
 && \beta \mapsto \beta + \zeta^{(s)\,\prime} - \zeta^{0} ~,
    \hspace{0,7cm}
    B_i \mapsto B_i + \zeta^{(v)\,\prime}_i ~,
    \hspace{0,5cm}
    C_i \mapsto C_i - \zeta^{(v)}_i ~,
    \label{e:gauge_trans5}
\end{eqnarray}
where we use the notation $~'\equiv d/d\eta$ and ${\cal H} \equiv
a'/a$. The spatial gauge-invariance of the shear, $\chi \equiv a
\left(\beta+\gamma'\right) \mapsto \chi - a\zeta^{0}$, enables us to
define an infinite number of gauge-invariant quantities.  For
instance, the Bardeen potentials are built exclusively from metric
variables
\begin{eqnarray*}
 \Phi \equiv \varphi - \frac{{\cal H}}{a}\chi~,
 \qquad \textrm{and} \qquad
 \Psi \equiv \tau-\frac{\chi'}{a}~.
 \label{e:pot-bardeen}
\end{eqnarray*}
Consider a four-scalar $\bar\phi(\eta,{\mathbf x}) = \phi(\eta) +
\delta\phi(\eta,{\mathbf x})$. Under the transformation laws
Eq.~(\ref{e:gauge_trans1}), the
perturbed part behaves as $\delta\phi \mapsto
\delta\phi-\phi'\zeta^{0}$. Hence, one may also construct
field-dependent gauge-invariant variables such as
\begin{eqnarray*}
 -{\cal R} = \varphi_{\delta\phi}
    \equiv \varphi-\frac{\cal H}{\phi'}\delta\phi
    \equiv -\frac{\cal H}{\phi'}\delta\phi_\varphi~,
 \qquad \textrm{or} \qquad
 \delta\phi_\chi
    \equiv \delta\phi-\frac{\phi'}{a}\chi
    \equiv -\frac{\phi'}{a}\chi_{\delta\phi}~.
 \label{e:gauge-inv-var}
\end{eqnarray*}
Choosing the uniform-field gauge $\delta \phi \equiv 0$ as the
temporal gauge condition promotes the perturbed variable $\varphi$ to
become a gauge-invariant quantity, which we denote by
$\varphi_{\delta\phi}$. Similarly, $-\frac{\cal H}{\phi'}
\delta\phi_\varphi$ becomes gauge-invariant in the uniform-curvature
gauge, $\varphi\equiv0$. Up to a sign, it is equal to the curvature
perturbation in the comoving gauge ${\cal R}$~\cite{Lyth:1985gv}. The
longitudinal gauge ($\beta=\gamma=0$, hence $\chi = 0$) renders the
field perturbation $\delta\phi$ gauge-invariant, which we denote by
$\delta\phi_\chi$. We then find
\begin{eqnarray*}
 \delta\phi_{\chi}
 = \frac{\phi'}{{\cal H}}
   \left[\Phi-\varphi_{\delta\phi}\right]
 = \frac{\phi'}{{\cal H}}
   \left[\Phi+{\cal R}\right]~.
 \label{e:conversion}
\end{eqnarray*}
If we are to consider the longitudinal gauge to fix the temporal
degree of freedom, the Bardeen potentials then coincide with the
perturbations in the metric, $\Phi = \varphi$ and $\Psi=\tau$,
while $\delta\phi_{\chi} = \delta\phi$.

\subsection{Background evolution}
\label{a:background}

If we neglect perturbations in the metric and field, the
infinitesimal line element reduces to $ds^2 = a^2(\eta)(-d\eta^2
+d\vec{x}^2)$. A solution for the background evolution is then
required to satisfy the Einstein equation and the dynamical equation for
the scalar field,
\begin{eqnarray}
 && G^\mu_\nu = \frac{1}{2{\cal F}}
    \left[{T^{(0)}}^\mu_\nu+\alpha'\lambda {T^{(c)}}^\mu_\nu\right]~,
    \label{e:app-Einstein}\\
 && \phi'' + 2{\cal H}\phi'- \frac{1}{2\omega} \Bigl[
 6{\cal F}_{,\phi} \left({\cal H}'+{\cal H}^2\right)
 -\omega_{,\phi}{\phi'}^2-2a^2{\cal V}_{,\phi}
 + \alpha'\lambda a^2 \Delta^{(c)}_\phi \Bigr] = 0~,
     \label{e:app-field}
\end{eqnarray}
where $~_{,\phi}\equiv d/d\phi$, and the different contributions are
\begin{eqnarray*}
 G^0_0 &=& -3a^{-2}{\cal H}^2~, \\
 G^i_j &=& -a^{-2}\left(2{\cal H}'+{\cal H}^2\right)\delta^i_j~,\\
 {T^{(0)}}^0_0 &=& a^{-2}\Big[6{\cal F}_{,\phi}{\cal H}
    \phi'-\omega\phi^{\prime 2}-2a^2{\cal V} \Big]~, \\
 {T^{(0)}}^i_j &=& a^{-2}\Big[2{\cal F}_{,\phi}\left(
    \phi''+{\cal H}\phi'\right)+\left(\omega+2{\cal F}_{,\phi\phi}
    \right)\phi^{\prime 2}-2a^2{\cal V}\Big]\delta^i_j~, \\
    && \nonumber \\
 {T^{(c_1)}}^0_0
    &=& 24 c_1 a^{-4} \xi_{,\phi}\phi' {\cal H}^3~,
    \label{e:G-T00-c1}\\
 {T^{(c_2)}}^0_0
    &=& -9 c_2 a^{-4}\xi\phi^{\prime 2}{\cal H}^2 ~,\\
 {T^{(c_3)}}^0_0
    &=& c_3 a^{-4} \phi^{\prime 3}
        \left(\xi_{,\phi}\phi'-6 \xi {\cal H} \right)~,\\
 {T^{(c_4)}}^0_0
    &=& -3 c_4 a^{-4}\xi \phi^{\prime 4} ~, \\
    && \nonumber \\
 {T^{(c_1)}}^i_j
    &=& 8 c_1 a^{-4}\Bigl[\left(\xi_{,\phi\phi}\phi^{\prime 2}
    +\xi_{,\phi}\phi''\right){\cal H}^2+\xi_{,\phi}\phi'{\cal H}
    \left(2{\cal H}'-{\cal H}^2\right)\Bigr]\delta^i_j~,\\
 {T^{(c_2)}}^i_j
    &=& -c_2 a^{-4}\phi'\Bigr[\xi\phi'
    \left(2{\cal H}'+{\cal H}^2\right)
    + 4 \xi \left(\phi''-\phi'{\cal H}\right){\cal H}
    + 2\xi_{,\phi}\phi^{\prime 2} {\cal H} \Bigr]\delta^i_j~,\\
 {T^{(c_3)}}^i_j
    &=& - c_3 a^{-4}\phi^{\prime 2}
    \Bigl[ 2\xi \left(\phi''-\phi'{\cal H}\right)
    +\xi_{,\phi}\phi^{\prime 2}\Bigr]\delta^i_j~,\\
 {T^{(c_4)}}^i_j
    &=&  c_4 a^{-4}\xi \phi^{\prime 4}\delta^i_j~,
    \label{e:G-Tij-c4} \\
 &&  \nonumber \\
 \Delta^{(c_1)}_{\phi}
   &=& 24c_1a^{-4}\xi_{,\phi}{\cal H}'{\cal H}^2~,
       \label{e:BGphi4d-1}\\
 \Delta^{(c_2)}_{\phi}
   &=& -3 c_2 a^{-4}{\cal H}\Big[
       \xi_{,\phi}\phi^{\prime 2}{\cal H}
       +2\xi\left(\phi''{\cal H}+2\phi'{\cal H}'\right)\Big] ~,\\
 \Delta^{(c_3)}_{\phi}
   &=& c_3 a^{-4} \phi'\Big[
       \xi_{,\phi\phi}\phi^{\prime 3}+4\xi_{,\phi}\phi'\left(
       \phi''-{\cal H}\phi'\right)-6\xi\Big(\phi'{\cal H}'
       +2\phi''{\cal H}\Big)\Big]~,\\
 \Delta^{(c_4)}_{\phi}
   &=& -3c_4 a^{-4}\phi^{\prime 2}\Big[
       \xi_{,\phi}\phi^{\prime 2}+4\xi\phi''\Bigr]~.
       \label{e:BGphi4d-4}
\end{eqnarray*}

\subsection{Linear perturbations}
\label{a:perturbation}

Assuming that there exists a class of background cosmologies which
satisfy Eqs.~(\ref{e:app-Einstein})--(\ref{e:app-field}) at all times,
the equations of motion for small perturbations linearised about the
background metric are
\begin{eqnarray}
 \delta G^\mu_\nu
 &=& \frac{1}{2{\cal F}}\left[\delta {T^{(0)}}^\mu_\nu
     + \alpha'\lambda \delta {T^{(c)}}^\mu_\nu
     - 2{\cal F}_{,\phi} G^\mu_\nu \delta\phi\right]~.
 \label{e:G-EOM-perturbations}
\end{eqnarray}
Following~\cite{Mukhanov:1992tc}, the purely geometrical part
obeys
\begin{eqnarray*}
 \delta G^0_0 &=& \left(\delta G^0_0\right)^{(GI)} +
   \left(G^0_0\right)'\frac{\chi}{a} ~,
   \label{e:G-G00-inv-def}\\
 \delta G^0_j &=& \left(\delta G^0_j\right)^{(GI)} +
   \left(G^0_0-\frac{1}{3}G^k_k\right)
   \left(\frac{\chi}{a}\right)_{|j}~,
   \label{e:G-G0j-inv-def}\\
 \delta G^i_j &=& \left(\delta G^i_j\right)^{(GI)} +
    \left(G^i_j\right)'\frac{\chi}{a}~,
    \label{e:G-Gij-inv-def}
\end{eqnarray*}
and so do the terms on the right hand side of
Eq.~(\ref{e:G-EOM-perturbations}). Using the background solution,
the gauge-invariant equations of motion for small perturbations
linearised about the background metric and field are
\begin{eqnarray*}
 \left(\delta {G}^0_0\right)^{(GI)}
 &=& \frac{1}{2{\cal F}}\left[\left(\delta {T^{(0)}}^0_0\right)^{(GI)}
     + \alpha'\lambda \left(\delta {T^{(c)}}^0_0\right)^{(GI)}
     - 2{\cal F}_{,\phi} G^0_0 \delta\phi_\chi\right]~,
     \label{e:G-G00-inv-pert} \\
 \left(\delta {G}^0_j\right)^{(GI)}
 &=& \frac{1}{2{\cal F}}\left[\left(\delta {T^{(0)}}^0_j\right)^{(GI)}
     + \alpha'\lambda \left(\delta {T^{(c)}}^0_j\right)^{(GI)}\right]~,
     \label{e:G-G0j-inv-pert} \\
 \left(\delta {G}^i_j\right)^{(GI)}
 &=& \frac{1}{2{\cal F}}\left[\left(\delta {T^{(0)}}^i_j\right)^{(GI)}
     + \alpha'\lambda \left(\delta {T^{(c)}}^i_j\right)^{(GI)}
     - 2{\cal F}_{,\phi} G^i_j \delta\phi_\chi\right]~,
     \label{e:G-Gij-inv-pert} \\
 \left(\delta {G}\right)^{(GI)}_{tf}
 &=& \frac{1}{2{\cal F}}\left[\left(\delta {T^{(0)}}\right)^{(GI)}_{tf}
     + \alpha'\lambda \left(\delta {T^{(c)}}\right)^{(GI)}_{tf}\right]~.
     \label{e:G-Gtf-inv-pert}
\end{eqnarray*}
Here, we defined $({\cal T})_{tf} \equiv {\cal T}^i_j - \frac{1}{3}{\cal
T}^k_k\delta^i_j$, i.e., the tracefree part of a tensorial quantity ${\cal T}^i_j$.
Then, using gauge-invariant quantities, the dynamical equation for
the scalar field $\phi$ reads
\begin{eqnarray*}
 0&=& 2\omega\delta\phi''_\chi +2\left(2{\cal H}\omega+\omega_{,\phi}\phi'
      \right)\delta\phi'_\chi \nonumber \\
   && +\Big[2\omega_{,\phi}\left(\phi''+2{\cal H}\phi'
      \right)-6{\cal F}_{,\phi\phi}\left({\cal H}'+{\cal H}^2\right)
      +\omega_{,\phi\phi}\phi^{\prime 2}+2a^2{\cal V}_{,\phi\phi}-2\omega
      \Delta\Big]\delta\phi_\chi\nonumber \\
   && -6{\cal F}_{,\phi}\Phi''-2\left(3{\cal F}_{,\phi}{\cal H}
      -\omega\phi'\right)\left(3\Phi'-\Psi'\right)+4a^2{\cal V}_{,\phi}\Psi
      +2{\cal F}_{,\phi}\Delta\left(2\Phi+\Psi\right)\nonumber \\
   && -\alpha'\lambda a^2 \left[2\Delta^{(c)}_\phi\Psi+
      \left(\delta \Delta^{(c)}_\phi\right)^{(GI)} \right]~,
 \label{e:G-equ-phi}
\end{eqnarray*}
where we used
\begin{eqnarray*}
 \delta \Delta^{(c_i)}_\phi =
 \left(\delta \Delta^{(c_i)}_\phi\right)^{(GI)} +
 \left(\Delta^{(c_i)}_\phi\right)'\frac{\chi}{a}~.
\end{eqnarray*}
The gauge-invariant components entering the above equations are
\begin{eqnarray*}
 \left(\delta {G}^0_0\right)^{(GI)} &=&
    -2a^{-2}\left[3{\cal H}\left(\Phi'
    -\Psi{\cal H}\right)-\Delta\Phi\right] ~,
    \label{e:G-G00-inv}\\
 \left(\delta {G}^0_j\right)^{(GI)} &=&
    2a^{-2}\left[\Phi'-\Psi{\cal H}\right]_{|j}~,
    \label{e:G-G0j-inv}\\
 \left(\delta {G}^i_j\right)^{(GI)} &=&
    -2a^{-2}\left[\Phi''+ \left(2\Phi'-\Psi'\right){\cal H}
    -\Psi\left(2{\cal H}'+{\cal H}^2\right)\right]\delta^i_j
    \nonumber \\ &&
    -a^{-2}\left(\nabla^i\nabla_j-\Delta\delta^i_j\right)
    \left(\Phi+\Psi\right)~,
    \label{e:G-Gij-inv}\\
    && \nonumber \\
  \left(\delta {T^{(0)}}^0_0\right)^{(GI)} &=&
    a^{-2}\Big[6{\cal F}_{,\phi\phi}{\cal H}\phi'
    -\omega_{,\phi}\phi^{\prime 2}-2a^2{\cal V}_{,\phi}
    -2{\cal F}_{,\phi}\Delta \Big]\delta\phi_\chi
    \nonumber \\ &&
    +2a^{-2}\Big[\left(3{\cal H}{\cal F}_{,\phi}
    -\omega\phi'\right)\delta\phi'_\chi+3{\cal F}_{,\phi}\phi'\Phi'
    -\phi'\left(6{\cal F}_{,\phi}{\cal H}-\omega\phi'\right)\Psi\Big]~,
    \label{e:G-T00-inv} \\
 \left(\delta {T^{(0)}}^0_j\right)^{(GI)} &=&
    -2a^{-2}\Big[{\cal F}_{,\phi}\delta\phi'_\chi
    +\left({\cal F}_{,\phi\phi}\phi'+\omega\phi'-{\cal F}_{,\phi}
    {\cal H}\right)\delta\phi_\chi-{\cal F}_{,\phi}\phi'\Psi
    \Big]_{\,|j}~,
    \label{e:G-T0j-inv}\\
 \left(\delta {T^{(0)}}^i_j\right)^{(GI)} &=&
    a^{-2}\Big\{2{\cal F}_{,\phi}\delta\phi''_\chi
    +2\Big[{\cal F}_{,\phi}{\cal H}+\phi'
    \left(\omega+2{\cal F}_{,\phi\phi}\right)\Big]\delta\phi'_\chi
    \nonumber \\ && \hspace{0,8cm}
    +\Big[2{\cal F}_{,\phi\phi}\left(\phi''+{\cal H}\phi'\right)+
    \left(\omega_{,\phi}+2{\cal F}_{,\phi\phi\phi}\right)\phi^{\prime 2}
    -\frac{2}{a^2}{\cal V}_{,\phi}\Big]\delta\phi_\chi
    \nonumber \\ && \hspace{0,8cm}
    +2{\cal F}_{,\phi}\phi'\left(2\Phi'-\Psi'\right)
    -2\Big[2{\cal F}_{,\phi}\left(\phi''+{\cal H}\phi'\right)
    +\phi^{\prime 2}\left(\omega+2{\cal F}_{,\phi\phi}\right)\Big]\Psi
    \Bigl\}\delta^i_j
    \nonumber \\ &&
    +2a^{-2}{\cal F}_{,\phi}
    \left(\nabla^i\nabla_j-\Delta\delta^i_j\right)\delta\phi_\chi~,
    \label{e:G-Tij-inv}\\
    && \nonumber \\
    && \nonumber \\
  \left(\delta {T^{(c_1)}}^0_0\right)^{(GI)} &=&
    8 c_1 a^{-4} {\cal H} \left[
    3\xi_{,\phi}{\cal H}^2\delta\phi'_{\chi}
    +3\xi_{,\phi\phi}{\cal H}^2\phi'\delta\phi_{\chi}
    +3\xi_{,\phi}{\cal H}\phi'\left(3\Phi'-4{\cal H}\Psi\right)\right.
    \nonumber \\ && \hspace{1,6cm}
    \left. -\xi_{,\phi}\Delta\left(2\phi'\Phi
    +{\cal H}\delta\phi_\chi\right)\right]~,
    \label{e:G-Tc100-inv} \\
 \left(\delta {T^{(c_1)}}^0_j\right)^{(GI)} &=&
    -8 c_1a^{-4}{\cal H}\left[
    \xi_{,\phi}{\cal H}\delta\phi'_\chi
    +{\cal H}\left(\xi_{,\phi\phi}\phi'-\xi_{,\phi}
    {\cal H}\right)\delta\phi_\chi
    +\xi_{,\phi}\phi'\left(2\Phi'-3{\cal H}\Psi\right) \right]_{\,|j}~,
    \label{e:G-Tc10j-inv}\\
 \left(\delta {T^{(c_1)}}^i_j\right)^{(GI)} &=&
    8c_1a^{-4}\bigg\{
    \xi_{,\phi}{\cal H}^2\delta\phi''_\chi
    + {\cal H}\left(2\xi_{,\phi\phi}{\cal H}\phi'+2\xi_{,\phi}{\cal H}'
    -\xi_{,\phi}{\cal H}^2\right)\delta\phi'_\chi
    \nonumber \\ && \hspace{1,4cm}
    +{\cal H}\left(\xi_{,\phi\phi\phi}{\cal H}\phi^{\prime 2}
    +\xi_{,\phi\phi}\left\{2{\cal H}'\phi'+{\cal H}\phi''
    -{\cal H}^2\phi'\right\}\right)\delta\phi_\chi
    \nonumber \\ && \hspace{1,4cm}
    + 2\xi_{,\phi}{\cal H}\phi'\Phi''
    +2\left(\xi_{,\phi\phi}{\cal H}\phi^{\prime 2} + \xi_{,\phi}
    {\cal H}'\phi'+\xi_{,\phi}{\cal H}\phi''\right)\Phi'
    -3\xi_{,\phi}{\cal H}^2\phi'\Psi'
    \nonumber \\ && \hspace{1,4cm}
    -4{\cal H}\left(\xi_{,\phi\phi}{\cal H}\phi^{\prime 2} + \xi_{,\phi}
    \left\{2{\cal H}'\phi'+{\cal H}\phi''-{\cal H}^2\phi'\right\}\right)
    \Psi\bigg\}\delta^i_j
    \\ &&
    +8c_1a^{-4}\left(\nabla^i\nabla_j-\Delta\delta^i_j\right)
    \Big[\xi_{,\phi}{\cal H}'\delta\phi_\chi
    +\left(\xi_{,\phi\phi}\phi^{\prime 2}+\xi_{,\phi}\left[\phi''
    -{\cal H}\phi'\right]\right)\Phi
    +\xi_{,\phi}{\cal H}\phi'\Psi\Big]~,
    \label{e:G-Tc1ij-inv}\nonumber \\
    && \nonumber \\
    && \nonumber \\
 \left(\delta {T^{(c_2)}}^0_0\right)^{(GI)} &=&
    -c_2 a^{-4}\phi'\left[18\xi{\cal H}^2\delta\phi_\chi'
    +9\xi_{,\phi}{\cal H}^2\phi'\delta\phi_\chi
    -4\xi{\cal H}\Delta\delta\phi_\chi \right.
    \nonumber \\ && \hspace{1,7cm}
    \left.+18\xi{\cal H}\phi'\left(\Phi'-2{\cal H}\Psi\right)
    -2\xi\phi'\Delta\Phi\right]~,
    \label{e:G-Tc200-inv} \\
 \left(\delta {T^{(c_2)}}^0_j\right)^{(GI)} &=&
    2c_2a^{-4}\phi'\left[2\xi{\cal H}\delta\phi_\chi'
    +{\cal H}\left(\xi_{,\phi}\phi'-3\xi{\cal H}\right)\delta\phi_\chi
    +\xi\phi'\left(\Phi'-3{\cal H}\Psi\right)\right]_{|j}~,
    \label{e:G-Tc20j-inv}\\
 \left(\delta {T^{(c_2)}}^i_j\right)^{(GI)} &=&
    -c_2a^{-4}\bigg\{
    4\xi{\cal H}\phi'\delta\phi_\chi''
    +2\left[\xi\left(2{\cal H}'\phi'+2{\cal H}\phi''-3{\cal H}^2\phi'\right)
    +3\xi_{,\phi}{\cal H}\phi^{\prime 2}\right]\delta\phi_\chi'
    \nonumber \\ && \hspace{1,45cm}
    +\phi'\left[2\xi_{,\phi\phi}{\cal H}\phi^{\prime 2}
    +\xi_{,\phi}\left(2{\cal H}'\phi'+4{\cal H}\phi''-3{\cal H}^2\phi'\right)
    \right]\delta\phi_\chi
    \nonumber \\ && \hspace{1,45cm}
    +2\xi\phi^{\prime 2}\Phi''
    +2\phi'\left(\xi_{,\phi}\phi^{\prime 2}+2\xi\phi''\right)\Phi'
    -6\xi{\cal H}\phi^{\prime 2}\Psi'
    \nonumber \\ && \hspace{1,45cm}
    -4\phi'\left[2\xi_{,\phi}{\cal H}\phi^{\prime 2}+\xi\left(
    2{\cal H}'\phi'+4{\cal H}\phi''-3{\cal H}^2\phi'\right)\right]\Psi
    \bigg\}\delta^i_j
    \nonumber \\ &&
    -c_2a^{-4}\left(\nabla^i\nabla_j-\Delta\delta^i_j\right)
    \left[\left(\xi_{,\phi}\phi^{\prime 2}+2\xi\phi''\right)\delta\phi_\chi
    -\xi\phi^{\prime 2}\left(\Phi-\Psi\right)\right]~,
    \label{e:G-Tc2ij-inv}\\
    && \nonumber \\
    && \nonumber \\
 \left(\delta {T^{(c_3)}}^0_0\right)^{(GI)} &=&
    c_3 a^{-4}\phi^{\prime 2}\bigg\{2\left(2\xi_{,\phi}\phi'-9\xi{\cal H}\right)
    \delta\phi_\chi'+\phi'\left(\xi_{,\phi\phi}\phi'-6\xi_{,\phi}{\cal H}
    \right)\delta\phi_\chi+2\xi\Delta\delta\phi_\chi
    \nonumber \\ && \hspace{1,65cm}
    -6\xi\phi'\Phi'-4\phi'\left(\xi_{,\phi}\phi'-6\xi{\cal H}\right)\Psi\bigg\}~,
    \label{e:G-Tc300-inv} \\
 \left(\delta {T^{(c_3)}}^0_j\right)^{(GI)} &=&
    2c_3 a^{-4}\phi'\left[\xi\phi'\delta\phi_\chi'+\phi'\left(\xi_{,\phi}\phi'
    -3\xi{\cal H}\right)\delta\phi_\chi-\xi\phi^{\prime 2}\Psi\right]_{|j}~,
    \label{e:G-Tc30j-inv}\\
 \left(\delta {T^{(c_3)}}^i_j\right)^{(GI)} &=&
    -c_3 a^{-4}\phi'\bigg\{2\xi\phi'\delta\phi_\chi''
    +2\left[2\xi_{,\phi}\phi^{\prime 2}+\xi\left(2\phi''
    -3{\cal H}\phi'\right)\right]\delta\phi_\chi'
    -2\xi\phi^{\prime 2}\Psi'
    \nonumber \\ && \hspace{1,7cm}
    +\phi'\left[\xi_{,\phi\phi}
    \phi^{\prime 2}+2\xi_{,\phi}\left(\phi''-{\cal H}\phi'\right)\right]
    \delta\phi_\chi
    \nonumber \\ && \hspace{1,7cm}
    -4\phi'\left[\xi_{,\phi}\phi^{\prime 2}
    +2\xi\left(\phi''-{\cal H}\phi'\right)\right]\Psi\bigg\}\delta^i_j~,
    \label{e:G-Tc3ij-inv}\\
    && \nonumber \\
    && \nonumber \\
 \left(\delta {T^{(c_4)}}^0_0\right)^{(GI)} &=&
    -3c_4a^{-4}\phi^{\prime 3}
    \left[4\xi\delta\phi'_{\chi}+\xi_{,\phi}\phi'\delta\phi_\chi
    -4\xi\phi'\Psi\right]~,
    \label{e:G-Tc400-inv} \\
 \left(\delta {T^{(c_4)}}^0_j\right)^{(GI)} &=&
    -4c_4a^{-4}\xi\phi^{\prime 3}{\delta\phi_\chi}_{|j} ~,
    \label{e:G-Tc40j-inv}\\
 \left(\delta {T^{(c_4)}}^i_j\right)^{(GI)} &=&
    c_4a^{-4}\phi^{\prime 3}
    \left[4\xi\delta\phi'_{\chi}+\xi_{,\phi}\phi'\delta\phi_\chi
    -4\xi\phi'\Psi\right]\delta^i_j~,
    \label{e:G-Tc4ij-inv}\\
    && \nonumber \\
    && \nonumber \\
 \left(\delta \Delta^{(c_1)}_\phi\right)^{(GI)} &=&
    8c_1a^{-4}\xi_{,\phi}\left[3{\cal H}^2\Phi''+3{\cal H}
    \left({\cal H}^2+2{\cal H}'\right)\Phi'-3{\cal H}^2\left({\cal H}\Psi'
    +4{\cal H}'\Psi\right)\right.
    \nonumber \\ && \hspace{1,7cm}\left.
    -\Delta\left(2{\cal H}'\Phi+{\cal H}^2\Psi\right)\right]
    +24c_1a^{-4}\xi_{,\phi\phi}{\cal H}^2{\cal H}'\delta\phi_{\chi}~,
    \label{e:G-Phi-c1-inv} \\
 \left(\delta \Delta^{(c_2)}_\phi\right)^{(GI)} &=&
    -c_2 a^{-4}\Big[6\xi{\cal H}^2\delta\phi_\chi'' + 6{\cal H}\left(\xi_{,\phi}
    \phi'{\cal H}+2\xi{\cal H}'\right)\delta\phi_\chi'
    -2\xi\left(2{\cal H}'+{\cal H}^2\right)\Delta\delta\phi_\chi
    \nonumber \\ && \hspace{1,35cm}
    +3{\cal H}\left\{\xi_{,\phi\phi}{\cal H}\phi^{\prime 2}+2\xi_{,\phi}
    \left({\cal H}\phi'' +2{\cal H}'\phi'\right)\right\}\delta\phi_\chi
    +12\xi{\cal H}\phi'\Phi''
    \nonumber \\ && \hspace{1,35cm}
    +6\left\{\xi_{,\phi}{\cal H}\phi^{\prime 2}
    +\xi\left(2{\cal H}\phi''+2{\cal H}'\phi'
    +3{\cal H}^2\phi'\right)\right\}\Phi'
    \nonumber \\ && \hspace{1,35cm}
    -2\left(\xi_{,\phi}\phi^{\prime 2}+2\xi\phi''\right)\Delta\Phi
    -18\xi{\cal H}^2\phi'\Psi'
    \nonumber \\ && \hspace{1,35cm}
    -12{\cal H}\left\{\xi_{,\phi}{\cal H}\phi^{\prime 2}+2\xi
    \left(2{\cal H}'\phi'+{\cal H}\phi''\right)\right\}\Psi
    -4\xi{\cal H}\phi'\Delta\Psi\Big]~,
    \label{e:G-Phi-c2-inv}\\
 \left(\delta \Delta^{(c_3)}_\phi\right)^{(GI)} &=&
    c_3 a^{-4} \Big[
    4\phi'\left(\xi_{,\phi}\phi'-3\xi{\cal H}\right)\delta\phi_\chi''
    -2\phi^{\prime 2}\left(2\xi_{,\phi}\phi'-9\xi{\cal H}\right)\Psi'
    + 2\xi\phi^{\prime 2}\Delta\Psi
    \nonumber \\ && \hspace{1,05cm}
    +4\left\{\xi_{,\phi\phi}\phi^{\prime 3}+\xi_{,\phi}\phi'
    \left(2\phi''-3{\cal H}\phi'\right)-3\xi \left({\cal H}'\phi'
    +{\cal H}\phi''\right)\right\}\delta\phi_\chi'
    \nonumber \\ && \hspace{1,05cm}
    +\phi'\left\{\xi_{,\phi\phi\phi}\phi^{\prime 3}+4\xi_{,\phi\phi}\phi'
    \left(\phi''-{\cal H}\phi'\right)-6\xi_{,\phi}\left(
    2{\cal H}\phi''+{\cal H}'\phi'\right)\right\}\delta\phi_\chi
    \nonumber \\ && \hspace{1,05cm}
    +4\xi\left(\phi''+{\cal H}\phi'\right)\Delta\delta\phi_\chi
    -6\xi\phi^{\prime 2}\Phi''-6\xi\phi'\left(2\phi''+3{\cal H}\phi'\right)\Phi'
    \nonumber \\ && \hspace{1,05cm}
    -4\phi'\left\{\xi_{,\phi\phi}\phi^{\prime 3}-4\xi_{,\phi}\phi'
    \left({\cal H}\phi'-\phi''\right)-6\xi\left({\cal H}'\phi'+2{\cal H}\phi''
    \right)\right\}\Psi\Big]~,
    \label{e:G-Phi-c3-inv}\\
 \left(\delta \Delta^{(c_4)}_\phi\right)^{(GI)} &=&
    -c_4a^{-4}\Big[12\xi\phi^{\prime 2}\delta\phi_\chi''
    +12\phi'\left(\xi_{,\phi}\phi^{\prime 2}+2\xi\phi''\right)\delta\phi_\chi'
    +3\phi^{\prime 2}\left(\xi_{,\phi\phi}\phi^{\prime 2}+4\xi_{,\phi}\phi''
    \right)\delta\phi_\chi
    \nonumber \\ && \hspace{1,35cm}
    -4\xi\phi^{\prime 2}\Delta\delta\phi_\chi
    +12\xi\phi^{\prime 3}\left(\Phi'-\Psi'\right)
    -12\phi^{\prime 2}\left(\xi_{,\phi}\phi^{\prime 2}
    +4\xi\phi''\right)\Psi\Big]~.
    \label{e:G-Phi-c4-inv}
\end{eqnarray*}

\subsection{Structure of the perturbation equations}
\label{ss:}

\begin{figure}[t]
 \begin{center}
 \includegraphics[width=0.45\linewidth]{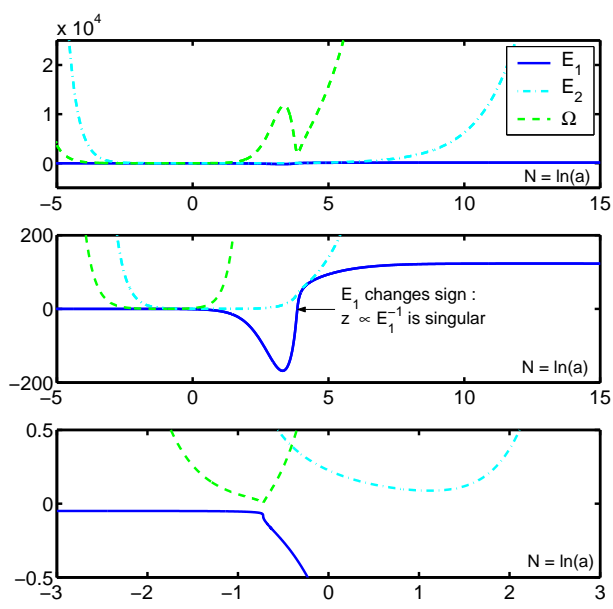}
 \caption{Here we illustrate the evolution of the quantities $E_{1}$,
 $E_{2}$ and $\Omega$ as functions of the number of e-folds, $N=\ln(a)$,
 in the context of the high-curvature transition of short duration
 used to determine the spectral distribution of Fig.~\ref{f:spectrum}.
 As explained in the main text, $E_{1}$ has to change sign
 in a bouncing universe. The pump field $z=\left(\phi'E_{2}/E_{1}^2\right)^{1/2}$ and
 $z'/z$ are thus singular, which invalidates the
 use of the homogeneous second-order differential equation for the
 curvature perturbation in the comoving gauge,
 ${\cal R}''+2\frac{z'}{z}{\cal R}' +s k^2 {\cal R}= 0$.
 Since $E_{2}$ and $\Omega$ do not change sign, the homogeneous
 systems of linear differential equations
 Eq.~(\ref{e:dyn-sys-Phi-dp}) and  Eq.~(\ref{e:dyn-sys-Phi-R})
 remain well-defined at all times, and can be
 used to determine the spectral distribution of adiabatic density perturbations
 long after the bounce.}
 \label{f:z-singular}
 \end{center}
\end{figure}

To visualise the structure of these complicated contributions, we write down
the perturbation equations in matrix form,
\begin{equation}
 {\cal M}^{(\Phi)}\hat\Phi+{\cal M}^{(\Psi)}\hat\Psi
 +{\cal M}^{(\delta\phi_\chi)}\hat{\delta\phi_\chi}=0  ~,
 \label{e:G-matrix-wave-gen}
\end{equation}
where we use $\hat{x}=\left(x'', x', \Delta x, x\right)^T$, $x \in
\{\Phi, \Psi,\delta\phi_\chi\}$. According to the gauge-invariant
contributions of Sec.~\ref{a:perturbation}, the ${\cal M}^{(x)}$
represent $5\times4$ matrices, each row of
Eq.~(\ref{e:G-matrix-wave-gen}) yielding one of the five component
perturbation equations. Here we choose that the first three rows
correspond to the $\left( ^0_0 \right)$, $\left( ^0_j \right)$ and
$\left( ^i_i \right)$ part of the perturbed Einstein equation, while
the fourth row correspond to the $\left( \phi \right)$ perturbation
equations. Finally, the tracefree part of the Einstein equation yields
the fifth row. Explicitly, after use of the background equations, the
matrices entering Eq.~(\ref{e:G-matrix-wave-gen}) have the following
structure
\begin{eqnarray*}
 &&{\cal M}^{(\Phi)} =
 \pmatrix{0 & -6 E_1 & 2 E_2 & 0 \cr 0 & 2 E_2 & 0 & 0 \cr
          -6 E_2 & -6 E_9 & 2 E_3 & 0 \cr  -6 E_4 & 6 E_8 & 2 E_6 & 0 \cr
          0 & 0 & 0 & -E_3}~,
 \hspace{2cm}{\cal M}^{(\Psi)} =
 \pmatrix{0 & 0 & 0 & E_{17} \cr 0 & 0 & 0 & -2 E_1 \cr
          0 & 6 E_1 & 2 E_2 & 3 E_{18} \cr 0 & -2 E_5 & 2 E_4 & 2 E_{19} \cr
          0 & 0 & 0 & - E_2}~, \nonumber \\
 &&{\cal M}^{(\delta\phi_\chi)} =
 \pmatrix{0 & 2 E_5 & 2 E_4 & - E_{10} \cr
          0 & 2 E_4 & 0 & 2 E_{11} \cr
          -6E_4 & -6 E_{20} & 2 E_6 & -3 E_{12} \cr
          2 E_{14} & 2 E_{15} & -2 E_{16}& E_{13} \cr 0 & 0 & 0 & -E_6}~,
\end{eqnarray*}
where we have highlighted the components which are null at all
times. It is then obvious that the tracefree part of the Einstein
equation, i.e., the fifth row of Eq.~(\ref{e:G-matrix-wave-gen}),
implies a linear relation among the perturbation variables
\begin{equation}
 \Psi = c_\Phi \Phi +c_{\delta\phi_\chi} \delta\phi_\chi~,
\end{equation}
where the coefficients $c_\Phi = -E_3 E_2^{-1}$ and
$c_{\delta\phi_\chi}=-E_6 E_2^{-1}$ are functions of the
background evolution, but do not depend explicitly on the
wavenumber $k$. Provided that the background evolution satisfies
$E_{2}\neq 0$ at all times, we can remove $\Psi$ from
Eq.~(\ref{e:G-matrix-wave-gen}), which yields a system of two
second--order differential equations and two constraints
$\tilde{{\cal M}}^{(\Phi)}\hat\Phi+\tilde{{\cal
M}}^{(\delta\phi_\chi)}\hat{\delta\phi_\chi}=0$, the $\tilde{{\cal
M}}^{(x)}$ corresponding to $4\times4$ matrices. It turns out that
some of these equations are redundant, and one can show that the
first two rows, i.e., the $\left( ^0_0 \right)$ and $\left( ^0_j
\right)$ components of the perturbed Einstein equation are
sufficient to fully characterise the dynamical evolution of
linear perturbations. Explicitly, we get
\begin{equation}
 \pmatrix{-6 E_1 & 2 E_2 & c_\Phi E_{17} \cr E_2 & 0 & -c_\Phi E_1 }
 \pmatrix{\Phi' \cr \Delta\Phi \cr \Phi}
 +
 \pmatrix{2 E_5 & 2 E_4 & -E_{10}+c_{\delta\phi_\chi}E_{17} \cr
          E_4 & 0 & E_{11}-c_{\delta\phi_\chi}E_1 }
  \pmatrix{\delta\phi_\chi' \cr \Delta\delta\phi_\chi \cr \delta\phi_\chi}
  = 0~.
\end{equation}
In Fourier space, these first-order equations can then be written
as ${\cal A}X' + {\cal B}X = 0$ where $X\equiv
(\Phi,\delta\phi_{\chi})^T$ and
\begin{equation}
 {\cal A}(\eta) = \pmatrix{-6E_1 & 2E_5 & \cr E_2 & E_4}~, \qquad
 {\cal B}(\eta,k) =
 \pmatrix{c_\Phi E_{17} - 2k^2 E_{2} & c_{\delta\phi_\chi}E_{17}-E_{10}-2k^2E_{4}\cr
          -c_\Phi E_{1} & E_{11}-c_{\delta\phi_\chi}E_1 }~.
\end{equation}
Provided that the matrix ${\cal A}$ is invertible, i.e.,
$\det({\cal A}) = -2\Omega \neq 0$ at all times, we then have $X'
= {\cal C}(\eta,k)X$ with ${\cal C}\equiv -{\cal A}^{-1}{\cal B}$.
Figure~\ref{f:z-singular} shows explicitly that $\Omega$ does not change sign
during the evolution for the background evolution discussed in the main text.
We can then proceed further by replacing the gauge-invariant field
perturbation, $\delta\phi_\chi = \frac{\phi'}{{\cal
H}}\big(\Phi+{\cal R}\big),$ which yields a system of first-order
coupled differential equations involving the Bardeen potential
$\Phi(\eta,k)$ and the curvature perturbation in the comoving
gauge ${\cal R}(\eta,k)$.

To be complete, we finally present the explicit form of the
coefficients of the matrices ${\cal M}^{(\Phi)}$, ${\cal
M}^{(\Psi)}$ and ${\cal M}^{(\delta\phi_\chi)}$. Not all of these
quantities are independent and one may use the background
equations (including the higher-order corrections) for additional
simplifications.

\begin{eqnarray*}
 E_1 &\equiv&  a^2\left({\cal F}'+2{\cal F}{\cal H}+a Q_{g}\right)~,
               \label{e:Element-1} \\
 E_2 &\equiv&  a^2\left(2{\cal F}+ Q_{b}\right)~,
               \label{e:Element-2} \\
 E_3 &\equiv&  a^2\left(2{\cal F}+ Q_{i}\right)~,
               \label{e:Element-3} \\
 E_4 &\equiv&  a^2\phi^{\prime -1}\left({\cal F}'+ a Q_{a}\right)~,
               \label{e:Element-4} \\
 E_5 &\equiv&  a^2\phi^{\prime -1}\left(\omega\phi^{\prime 2}
               -3{\cal F}'{\cal H}+ a^2 Q_{n}\right)~,
               \label{e:Element-5} \\
 E_6 &\equiv&  a^2\phi^{\prime -1}\left(2{\cal F}'+ a Q_{j}\right)~,
               \label{e:Element-6} \\
 E_7 &\equiv&  a^2\left(2{\cal F}+ Q_{m}\right)~,
               \label{e:Element-7} \\
 E_8 &\equiv&  a^2\phi^{\prime -1}\left(\omega\phi^{\prime 2}
               -3{\cal F}'{\cal H}+ a^2 Q_{l}\right)~,
               \label{e:Element-8} \\
 E_9 &\equiv&  2a^2\left({\cal F}'+2{\cal H}{\cal F}+a Q_{k}\right)~,
               \label{e:Element-9} \\
 E_{10} &\equiv& a^2\phi^{\prime -2}\left(
               6{\cal H}\left[{\cal F}''\phi'-{\cal F}'\phi''\right]
               +6{\cal F}'{\cal H}^2\phi'-\omega'\phi^{\prime 3}
              -2a^2{\cal V}'\phi'+a^4 Q_{s}\right)  ~,
               \label{e:Element-10} \\
 E_{11} &\equiv&  a^2\phi^{\prime -2}\left({\cal F}''\phi'-{\cal F}'\phi''
               -{\cal F}'{\cal H}\phi'+\omega\phi^{\prime 3}+a^3 Q_{t}\right)~,
               \label{e:Element-11} \\
 E_{12} &\equiv& a^2\phi^{\prime -3}\left(2{\cal F}'''\phi^{\prime 2}
            +2{\cal F}''\left[{\cal H}\phi'-2\phi''\right]\phi'
            +\omega'\phi^{\prime 4}-2a^2{\cal V}'\phi^{\prime 2}\right.
            \nonumber \\ && \hspace{1,2cm} \left.
            +2{\cal F}'\left[{\cal H}^2\phi^{\prime 2}
            +2{\cal H}'\phi^{\prime 2}-\phi'\phi'''
            +2\phi^{\prime \prime 2}-{\cal H}\phi'\phi''\right]
             +a^5 Q_{u}\right) ~,
               \label{e:Element-12} \\
 E_{13} &\equiv& a^2\phi^{\prime -3}\left(
            -6\left[{\cal F}''\phi'-{\cal F}'\phi''\right]
            \left[{\cal H}'+{\cal H}^2\right]
            +\omega''\phi^{\prime 3}+\omega'\phi^{\prime 2}
            \left[4{\cal H}\phi'+\phi''\right]\right.
            \nonumber \\ && \hspace{1,2cm} \left.
            +2a^2{\cal V}''\phi'-2a^2{\cal V}'\phi''+a^5 Q_{v}\right) ~,
               \label{e:Element-13} \\
 E_{14} &\equiv& a^2\phi^{\prime -2}\left(\omega\phi^{\prime 2}+a^2Q_{c}\right)~,
               \label{e:Element-14} \\
 E_{15} &\equiv&  a^2\left(\omega'+2{\cal H}\omega+a Q_{w}\right)~,
               \label{e:Element-15} \\
 E_{16} &\equiv& a^2\phi^{\prime -2}\left(\omega\phi^{\prime 2}+a^2Q_{o}\right)~,
               \label{e:Element-16} \\
 E_{17} &\equiv& 2a^2\left(6{\cal F}'{\cal H}+6{\cal F}{\cal H}^2
            -\omega\phi^{\prime 2}+a^2 Q_{p}\right)~,
               \label{e:Element-17} \\
 E_{18} &\equiv& 2a^2\left[2{\cal F}''+2{\cal F}'{\cal H}
            +2{\cal F}\left(2{\cal H}'+{\cal H}^2\right)
            +\omega\phi^{\prime 2}+a^2 Q_{q}\right]~,
               \label{e:Element-18} \\
 E_{19} &\equiv& a^4 \phi^{\prime -1} \left(2V'+a Q_{r}\right)~,
               \label{e:Element-19} \\
 E_{20} &\equiv& a^2\phi^{\prime -2}\left[2\left({\cal F}''\phi'-{\cal F}''\phi''\right)
            +{\cal F}'{\cal H}\phi'+\omega\phi^{\prime 3}+a^3 Q_{x}\right]~,
               \label{e:Element-20} \\
 && \nonumber \\
 Q_{a} &\equiv& \alpha'\lambda a^{-3}\left[4c_1\xi'{\cal H}^2
        -2c_2\xi{\cal H}\phi^{\prime 2}-c_3\xi\phi^{\prime 3}\right]~,
        \label{e:Q_scalar-a} \\
 Q_{b} &\equiv& \alpha'\lambda a^{-2}\left[8c_1\xi'{\cal H}
        -c_2\xi\phi^{\prime 2}\right]~,
        \label{e:Q_scalar-b} \\
 Q_{c} &\equiv& \alpha'\lambda a^{-4}\phi^{\prime 2}\left[
        3c_2\xi {\cal H}^2+2c_3\phi'\left(3\xi{\cal H}-\xi'\right)
        +6c_4\xi\phi^{\prime 2}\right]~,
        \label{e:Q_scalar-c} \\
 Q_{d} &\equiv& 2\alpha'\lambda a^{-4}\phi^{\prime 2}\left[
        c_2\xi\left({\cal H}'-{\cal H}^2\right)
        +c_3\left(\xi'\phi'+\xi\phi''-2\xi{\cal H}\phi'\right)
        -2c_4\xi\phi^{\prime 2}\right]~,
        \label{e:Q_scalar-d} \\
 Q_{e} &\equiv& 2\alpha'\lambda a^{-3}
        \left[8c_1\xi'\left({\cal H}'-{\cal H}^2\right)
        -c_2\phi'\left(\xi'\phi'+2\xi\phi''-4\xi\phi'{\cal H}\right)
        +2c_3\xi\phi^{\prime 3}\right]~,
        \label{e:Q_scalar-e} \\
 Q_{f} &\equiv& -2\alpha'\lambda a^{-2}\left[4c_1\left(
        \xi''-2\xi'{\cal H}\right)+c_2\xi\phi^{\prime 2}\right]~,
        \label{e:Q_scalar-f} \\
 Q_{g} &\equiv& \alpha'\lambda a^{-3}\left[12c_1\xi'{\cal H}^2
        -3c_2\xi{\cal H}\phi^{\prime 2}-c_3\xi\phi^{\prime 3}\right]~,
        \label{e:Q_scalar-g}\\
 Q_{h} &\equiv& 4\alpha'\lambda a^{-2} c_1 \xi''~,
        \label{e:Q_scalar-h}\\
 Q_{i} &\equiv& \alpha'\lambda a^{-2}\left[8c_1\left(\xi''-\xi'{\cal H}\right)
        +c_2\xi\phi^{\prime 2}\right]~,
        \label{e:Q_scalar-i}\\
 Q_{j} &\equiv& \alpha'\lambda a^{-3}\left[8c_1\xi'{\cal H}'
        -c_2\phi'\left(\xi'\phi'+2\xi\phi''\right)\right]~,
        \label{e:Q_scalar-j}\\
 Q_{k} &\equiv& \frac{1}{2}\alpha'\lambda a^{-3}
        \left[8c_1\left(\xi''{\cal H}+\xi'{\cal H}'\right)
        -c_2\phi'\left(\xi'\phi'+2\xi\phi''\right)\right] ~,
        \label{e:Q_scalar-k}\\
 Q_{l} &\equiv& \alpha'\lambda a^{-4}
        \left[-4c_1\xi'{\cal H}\left(2{\cal H}'+{\cal H}^2\right)
        +c_2\phi'\left(\xi'{\cal H}\phi'+\xi
        \left\{2{\cal H}'\phi'+2{\cal H}\phi''+3{\cal H}^2\phi'\right\}
        \right)\right. \nonumber \\ && \hspace{1,3cm}
        \left.+c_3\xi\phi^{\prime 2}\left(2\phi''+3{\cal H}\phi'\right)
        +2c_4\xi\phi^{\prime 4}\right]~,
        \label{e:Q_scalar-l}\\
 Q_{m} &\equiv& 8\alpha'\lambda c_1 a^{-2}\xi'{\cal H} ~,
        \label{e:Q_scalar-m} \\
 Q_{n} &\equiv& \alpha'\lambda a^{-4}
        \left[-12c_1\xi'{\cal H}^3+9c_2\xi{\cal H}^2\phi^{\prime 2}
        -c_3\phi^{\prime 3}\left(2\xi'-9\xi{\cal H}\right)
        +6c_4\xi\phi^{\prime 4}\right]~,
        \label{e:Q_scalar-n} \\
 Q_{o} &\equiv& \alpha'\lambda a^{-4}\phi^{\prime 2}\left[
        c_2\xi\left(2{\cal H}'+{\cal H}^2\right)+2c_3\xi\left(\phi''
        +{\cal H}\phi'\right)+2c_4\xi\phi^{\prime 2}\right]~,
        \label{e:Q_scalar-o} \\
 Q_{p} &\equiv& 2\alpha'\lambda a^{-4}\left[
        24c_1\xi'{\cal H}^3-9c_2\xi{\cal H}^2\phi^{\prime 2}
        +c_3\phi^{\prime 3}\left(\xi'-6\xi{\cal H}\right)
        -3c_4\xi\phi^{\prime 4}\right] ~,
        \label{e:Q_scalar-p} \\
 Q_{q} &\equiv& 2\alpha'\lambda a^{-4} \left[
        8c_1{\cal H}\left(\xi''{\cal H}+2\xi'{\cal H}'-\xi'{\cal H}^2\right)
        -c_2\phi'\left(2\xi'{\cal H}\phi'+2\xi{\cal H}'\phi'
        +4\xi{\cal H}\phi''-3\xi{\cal H}^2\phi'\right)\right.
        \nonumber \\ && \hspace{1,5cm}
        \left.-c_3\phi^{\prime 2}\left(\xi'\phi'+2\xi\phi''-2\xi{\cal H}\phi'
        \right)+c_4\xi\phi^{\prime 4}\right] ~,
        \label{e:Q_scalar-q} \\
 Q_{r} &\equiv& \alpha'\lambda a^{-5}\left[
        24c_1\xi'{\cal H}'{\cal H}^2-3c_2{\cal H}\phi'\left(
        \xi'{\cal H}\phi'+2\xi{\cal H}\phi''+4\xi{\cal H}'\phi'\right)\right.
        \nonumber \\ && \hspace{1,3cm}\left.
        +c_3\phi^{\prime 2}\left(\xi''\phi'+\xi'\left\{3\phi''-4{\cal H}\phi'
        \right\}-6\xi\left\{{\cal H}'\phi'+2{\cal H}\phi''\right\}\right)
        -3c_4\phi^{\prime 3}\left(\xi'\phi'+4\xi\phi''\right)\right]
        ~,\label{e:Q_scalar-r}\\
 Q_{s} &\equiv& \alpha'\lambda a^{-6}\left[
        24c_1\left(\xi''\phi'-\xi'\phi''\right){\cal H}^3
        -9c_2\xi'{\cal H}^2\phi^{\prime 3}
        +c_3\phi^{\prime 3}\left(\xi''\phi'-\xi'\phi''-6\xi'{\cal H}\phi'\right)
        -3c_4\xi'\phi^{\prime 5}
        \right]~,\label{e:Q_scalar-s}\\
 Q_{t} &\equiv& \alpha'\lambda a^{-5}\left[
        4c_1{\cal H}^2\left(\xi''\phi'-\xi'\phi''-\xi'{\cal H}\phi'\right)
        -c_2{\cal H}\phi^{\prime 3}\left(\xi'-3\xi{\cal H}\right)
        -c_3\phi^{\prime 4}\left(\xi'-3\xi{\cal H}\right)
        +2c_4\xi\phi^{\prime 5}
        \right]~,\label{e:Q_scalar-t}\\
 Q_{u} &\equiv& \alpha'\lambda a^{-7}\left[
        8c_1{\cal H}\left(\xi'''{\cal H}\phi^{\prime 2}
        +\xi''\phi'\left\{2{\cal H}'\phi'-{\cal H}^2\phi'-2{\cal H}\phi''\right\}
        -\xi'\left\{{\cal H}\phi'\phi'''+2{\cal H}'\phi'\phi''
        -2{\cal H}\phi^{\prime \prime 2}-{\cal H}^2\phi'\phi''\right\}\right)
        \right. \nonumber \\ && \hspace{1,25cm}\left.
        -c_2\phi^{\prime 3}\left(2\xi''{\cal H}\phi'+
        \xi'\left\{2{\cal H}'\phi'
        +2{\cal H}\phi''-3{\cal H}^2\phi'\right\}\right)-c_3\phi^{\prime 4}
        \left(\xi''\phi'+\xi'\phi''-2\xi'{\cal H}\phi'\right)
        +c_4\xi'\phi^{\prime 6}
        \right]~,\label{e:Q_scalar-u}\\
 Q_{v} &\equiv& \alpha'\lambda a^{-7}\left[
        -24 c_1{\cal H}'{\cal H}^2\left(\xi''\phi'-\xi'\phi''\right)
        +3c_2{\cal H}\phi^{\prime 2}
        \left(\xi''{\cal H}\phi'+\xi'{\cal H}\phi''+4\xi'{\cal H}'\phi'\right)
        \right. \nonumber \\ && \hspace{1,25cm}
        +c_3\phi^{\prime 2}\left(-\xi''' \phi^{\prime 2}+\xi''\phi'
        \left\{ 4{\cal H}\phi'-\phi''\right\}
        +\xi'\left\{ 8{\cal H}\phi'\phi''+\phi^{\prime\prime 2}+
        \phi'\phi'''+6{\cal H}'\phi^{\prime 2} \right\}\right)
        \nonumber \\ && \hspace{1,25cm}\left.
        +3c_4\phi^{\prime 4}\left(\xi''\phi'+3\xi'\phi''\right)
        \right]~,\label{e:Q_scalar-v}\\
 Q_{w} &\equiv& \alpha'\lambda a^{-3}\left[
        3c_2{\cal H}\left(\xi'{\cal H}+2\xi{\cal H}'\right)
        +6c_4\phi'\left(\xi'\phi'+2\xi\phi''\right)
        \right. \nonumber \\ && \hspace{1,25cm}\left.
        +2c_3\left(3\xi{\cal H}'\phi'-\xi'\phi''-\xi''\phi'
        +3\xi{\cal H}\phi''+3\xi'{\cal H}\phi'\right)
        \right]~,\label{e:Q_scalar-w}\\
 Q_{x} &\equiv& \alpha'\lambda a^{-5}\left[
        4c_1{\cal H}\left(2\xi''{\cal H}\phi'-2\xi'{\cal H}\phi''
        +2\xi'{\cal H}'\phi'-\xi'{\cal H}^2\phi'\right)
        +c_3\phi^{\prime 3}\left(-2\xi'\phi'+3\xi{\cal H}\phi'
        -2\xi\phi''\right)\right. \nonumber \\
       && \hspace{1,3cm}\left.
        -c_2\phi^{\prime 2}\left(3\xi'{\cal H}\phi'
        +2\xi{\cal H}\phi''+2\xi{\cal H}'\phi'
        -3\xi{\cal H}^2\phi'\right)+2c_4\xi\phi^{\prime 5}
        \right]~.\label{e:Q_scalar-x}
\end{eqnarray*}

\end{document}